\documentclass[usenatbib, epsfig, a4paper]{mn2e} 
\pdfoutput=1 

\usepackage[T1]{fontenc}
\usepackage{aecompl}
\usepackage{times}
\usepackage{subfigure}
\usepackage{amsmath,accents}
\usepackage{amsfonts}
\usepackage[pdftex]{graphicx}
\usepackage{multirow}
\usepackage{hyperref}
\usepackage{url}
\usepackage{bm}
\usepackage{ulem} 
\usepackage[abs]{overpic} 
\usepackage{epstopdf}  

\renewcommand{\d}[1]{\ensuremath{\operatorname{d}\!{#1}}}
\newcommand {\apgt} {\ {\raise-.5ex\hbox{$\buildrel>\over\sim$}}\ }
\newcommand {\aplt} {\ {\raise-.5ex\hbox{$\buildrel<\over\sim$}}\ }

\newenvironment{packed_enum}{
\begin{enumerate}
  \setlength{\itemsep}{1pt}
  \setlength{\parskip}{0pt}
  \setlength{\parsep}{0pt}
}{\end{enumerate}}

\newenvironment{packed_item}{
\begin{itemize}
  \setlength{\itemsep}{1pt}
  \setlength{\parskip}{0pt}
  \setlength{\parsep}{0pt}
}{\end{itemize}}

\def\myputfigure#1#2#3#4#5%
{\vskip#5pt\makebox[0pt]{\hskip#2in
\includegraphics[width=#3\textwidth]{#1}}\vskip#4pt\hfill}

\newcommand{\cmfast}{\textsc{\small 21CMFAST }}

\newcommand{\hi}{H\thinspace\textsc{i} }

\title[The impact of spin temperature fluctuations on 21-cm moments]{The impact of spin temperature fluctuations on the 21-cm moments}
\author[C. A. Watkinson \& J. R. Pritchard]
{C.~A.~Watkinson$^1$\thanks{Email: \href{mailto:c.watkinson11@imperial.ac.uk}{\protect\nolinkurl{c.watkinson11@imperial.ac.uk}}}\& J.~R.~Pritchard$^1$ \\ 
$^1$Department of Physics, Blackett Laboratory, Imperial College, London SW7 2AZ, UK\\
}
 
\date{\today}

\pubyear{\number\year}

\begin{document}
\maketitle

\begin{abstract}
This paper considers the impact of Lyman-$\alpha$
coupling and X-ray heating on the 21-cm brightness-temperature
one-point statistics (as predicted by semi-numerical
simulations).
The X-ray production efficiency is varied over
four orders of magnitude and the
hardness of the X-ray spectrum is varied from
that predicted for high-mass X-ray binaries,
to the softer spectrum expected from the hot inter-stellar medium.
We find peaks in the
redshift evolution of both the variance and skewness associated
with the efficiency of X-ray production.
The amplitude of the variance
is also sensitive to the hardness of the X-ray SED.
We find that the relative timing of
the coupling and heating phases can be inferred
from the redshift extent of a plateau that
connects a peak in the variance's
evolution associated with Lyman-$\alpha$ coupling to the heating peak.
Importantly,
we find that late X-ray heating would seriously
hamper our ability to
constrain reionization with the variance.
Late X-ray heating also qualitatively alters
the evolution of the skewness, providing a
clean way to constrain such models.
If foregrounds can be removed, we find that LOFAR, MWA and PAPER
could constrain reionization and late X-ray heating models with the variance.
We find that HERA and SKA (phase 1) will be able to constrain both
reionization and heating by measuring the variance using foreground-avoidance techniques.
If foregrounds can be removed they will also be able to constrain the nature of Lyman-$\alpha$ coupling.
\end{abstract}
\begin{keywords}
dark ages, reionization, first stars -- intergalactic medium -- methods: statistical -- cosmology: theory.
\end{keywords}

\section{Introduction}\label{sec:intro}
The cosmic dark ages, during which the only source of radiation
was the adiabatically cooling cosmic microwave
background (CMB),
ended when the first stars formed (see \citealt{Loeb2013} for an overview of this process).
The exact nature of the first stars and galaxies 
is uncertain, but the radiation
they emitted will have dramatically altered the state of
the intergalactic medium (IGM). Neutral 
hydrogen (H\thinspace\textsc{i}) dominates the composition
of the IGM until the epoch of reionization (EoR);
as such it is hoped that the impact of these galaxies will
be detectable with the 21-cm line, a spectral line
produced by a hyperfine transition in H\thinspace\textsc{i} 
\citep{Field1958, Field1959a}. 
The spin temperature ($T_{\rm s}$),
defines the distribution of the electron
population over the singlet and triplet 
hyperfine levels involved in 
the 21-cm transition \citep{EWEN1951}. 
If this is in equilibrium with the temperature
of the CMB ($T_{\textsc{cmb}}$) then 
the 21-cm signal will not be detectable.
However, radiation from stars breaks this
equilibrium, leading to an observable signal
in absorption where $T_{\rm s}<T_{\textsc{cmb}}$, and 
in emission when $T_{\rm s}>T_{\textsc{cmb}}$.

\subsection{The impact of the first stars}
As well as affecting the spin temperature,
radiation from the first stars
began ionizing neutral hydrogen.
Most important to this discussion is
the production of Lyman-$\alpha$,
X-ray and ultra-violet radiation.

\begin{packed_enum}
\item \textbf{Wouthuysen-Field (Lyman-$\alpha$) coupling:}. 
The populations of the 21-cm hyperfine levels
are mixed by repeated absorption and re-emission
of Lyman-$\alpha$ radiation.\footnote{Lyman-$n$ photons
contribute to the Lyman-$\alpha$ radiation field through cascades.} 
This couples the spin temperature to the Lyman-$\alpha$
colour temperature $T_\alpha$.
Repeated scattering of Lyman-$\alpha$ photons off 
atoms couples $T_\alpha$ 
to the kinetic gas temperature ($T_{\rm k}$)
and so $T_{\rm s}\sim T_{\rm k}$
\citep{Wouthuysen1952, Field1958, Pritchard2006}.
WF coupling therefore produces fluctuations
in the 21-cm signal, the observation of
which would provide insight 
into the nature of Lyman-$\alpha$ sources.
\item \textbf{X-ray heating:} An abundance of X-rays are produced
by accretion onto compact objects,
such as black holes and neutron stars,
as well as by hot gas in the interstellar medium. 
These X-rays induce photo-ionizations 
resulting in primary and secondary 
electrons. 
It is unlikely that X-ray photo-ionizations 
are efficient enough
to be solely responsible for the reionization of the universe
\citep{Dijkstra2004, Hickox2007, McQuinn2012};
however, once the IGM has
been ionized to a few percent, 
the photo-ejected electrons deposit the
majority ($\sim 65$\%) of their energy 
as heat in the IGM  
\citep{Shull1979, Shull1985a, Furlanetto2010}.
Because the Wouthuysen-Field (WF) coupling 
described in (i) 
is likely to have started 
at an early stage,
the onset of X-ray production will raise the 21-cm
spin temperature. Observation of 21-cm fluctuations 
produced by the heating process
will therefore provide insight into the nature 
of X-ray sources.

\item \textbf{Reionization:} 
As the Universe evolves, Ultra-Violet (UV) radiation from 
a growing number of galaxies begins to
ionize the IGM. These ionizing photons have
a short mean-free path and so carve out
well defined ionized regions around galaxies 
in an otherwise mostly neutral IGM. 
These grow with time until they
eventually merge and the Universe is 
reionized.
The associated depletion of 
neutral gas (and thus 21-cm signal)
produces 21-cm fluctuations whose observation
will provide insight into the process of reionization
(i.e. the nature of the sources responsible, and of the IGM)
Whilst sub-dominant to UV ionizations, 
X-ray induced ionizations will also impact 
on the ionization state of the IGM 
(for example, \citealt{PengOh2001, Venkatesan2001} and 
\citealt{Mesinger2013a}).
\end{packed_enum} 

The exact timing of processes
(i)-(iii), and the degree to which they overlap,
are uncertain, especially
for the coupling and heating processes.
This uncertainty depends on the nature of
the stars that drive WF coupling 
and the efficiency (and relative timing) 
of X-ray production \citep{Mesinger2013a}.
If the X-ray efficiency
is low enough, 21-cm fluctuations 
induced by X-ray heating may well persist
into the EoR.

Despite such uncertainties, 
it is expected that processes (i) - (iii)
occurred in the order described.
As the main source of X-rays is thought to be
accretion onto compact objects,
the production of X-rays is likely to be
delayed by a few million years relative to 
the first stars igniting.
Lyman-$\alpha$ production is coincident with
formation of the first stars, and
the emissivity to achieve Lyman-$\alpha$ 
coupling is much lower than that
required of X-rays to substantially heat the IGM.
Therefore Lyman-$\alpha$ coupling 
will at least commence prior
to the onset of X-ray heating \citep{Chen2004}. 
Because the mean free path of
UV photons is very small, UV-driven 
reionization will inevitably
be delayed relative to 
WF coupling and X-ray heating. 

\subsection{Constraints on X-ray production from post-EoR redshifts}\label{sec:conts}

The only constraints we have on the
nature of high-redshift X-ray production
come from lower-redshift observations. 
The dominant X-ray sources observed are
active galactic nuclei (AGN) and 
high-mass X-ray binaries (HMXBs).
X-ray emission from the hot interstellar medium (ISM) 
is also found to contribute significantly
to the soft X-ray emission of nearby galaxies
(e.g. \citealt{Mineo2012}).

Observations of the unresolved cosmic X-ray background 
point towards AGN being the dominant contributor 
in the local universe \citep{Moretti2012}.
However, the AGN number density rapidly decreases
at $z>3$ (see \citealt{Fan2001} and references therein),
although some scope remains in
the faint end of the luminosity function
for low-mass `mini-quasars'
to contribute at higher redshifts
(for example, \citealt{Madau2004, Volonteri2009}).

HMXBs are expected to be dominant at high redshift
because: (1) in the absence of an AGN they dominate X-ray 
output in low-redshift galaxies \citep{Fabbiano2006}, 
and (2) their abundance is proportional to 
star-formation rate and `starburst' galaxies
are seen to increase with redshift 
(for example, \citealt{Gilfanov2004, Mirabel2011, Mineo2012a}).
Theoretical modelling also suggests 
that a high fraction of the first stars
formed in binaries or multiple
systems would evolve into HMXBs 
\citep{Turk2009, Stacy2010}.

The X-ray spectral energy distribution (SED) 
and the associated mean free path of X-rays
determine the scale of 
21-cm fluctuations produced by inhomogeneous heating. 
The SED can be fit with a power law where the 
specific luminosity $L_{\mathrm X}$ is proportional 
to the frequency $\nu$ as 
$L_{\mathrm X} \propto \nu^{-\alpha}$ with
spectral energy index\footnote
{This is related to the photon index as $\Gamma = \alpha + 1$.} of $\alpha$. 
A harder X-ray SED, with $\alpha<1$,
produces higher energy X-rays,
which have a 
longer mean free path. 
Shocks in supernovae and AGN have an energy index $\alpha\sim 1.7$ 
\citep{Tozzi2006, McQuinn2012}, 
the hot ISM produces an SED described 
by $\alpha\sim 3$ \citep{Pacucci2014a},
and HMXBs have a hard 
spectra described by $\alpha\sim 0.7$ - $1$ 
\citep{Rephaeli1995, Swartz2004, Mineo2012a}.
Even if we knew the appropriate spectral energy index
to describe the SED at high redshift,
there is the matter of the 
luminosity's normalisation. 
We can normalise the luminosity at high redshifts 
to match the low redshift observations,
however it could be
orders of magnitudes 
apart from what we observe today.

\subsection{Observing and understanding the 21-cm signal}
Given that the X-ray properties 
of high redshift sources are so uncertain 
there is much to be gained from observing 
these epochs. The first generation of 
21-cm radio telescope such as 
LOFAR\footnote{The LOw Frequency ARray \url{http://www.lofar.org/}}, 
MWA\footnote{The Murchison Wide-field Array 
\url{http://www.mwatelescope.org/}} and 
PAPER\footnote{The Precision Array to Probe Epoch of Reionization 
\url{http://eor.berkeley.edu/}} 
aim to constrain reionization statistically
and are already starting to set interesting upper
limits (see \citealt{Paciga2011, Dillon2013, Ali2015} and \citealt{Pober2015}).
However, next-generation instruments 
HERA\footnote{The Hydrogen Epoch of Reionization Array 
\url{http://reionization.org/}} 
and 
SKA\footnote{The Square Kilometre Array 
\url{http://www.skatelescope.org/}} 
will not only dramatically improve
constraints on reionization, but
also aim to probe the pre-reionization era.
Telescopes seeking to measure the global average of
the 21-cm signal, such as EDGES\footnote{The Experiment to Detect the Global EoR Signal
\url{http://www.haystack.mit.edu/ast/arrays/Edges/}}
 and DARE\footnote{The Dark Ages Radio Explorer \url{http://lunar.colorado.edu/dare/}},
should also provide
valuable (and complementary) constraints on the EoR and
pre-reionization epochs \citep{bowman2010, Burns2012}.

Observing the 21-cm line will clearly be 
rewarding, however it will be challenging 
as the signal is small ($\sim 10$ mK) and 
foregrounds will be orders of magnitude larger 
\citep{Shaver1999a, DiMatteo2002a, PengOh2003a, DiMatteo2004a}.
It is hoped that by exploiting the spectral smoothness 
of foregrounds they may be removed (e.g. \citealt{Wang2006a, Liu2011a, Paciga2011, Petrovic2011b, Chapman2012, Cho2012, Shaw2014}). 
Alternatively, we could avoid foregrounds by
exploiting the existence of a wedge feature 
in the $k_\perp$-$k_\parallel$ cylindrically-binned
2D power spectrum to which foregrounds are confined 
(e.g. \citealt{Datta2010a, Vedantham2012, Morales2012, 
Thyagarajan2013, Hazelton2013, Liu2014}).\footnote{The
wedge results from the spectral smoothness of the foregrounds 
(which one might only expect to observe at low $k_\perp$)
combined with the chromatic nature of a radio telescope
(i.e. at different frequencies the instrument probes
different scales) which smear foregrounds 
into larger $k_\parallel$.}
It is not yet clear how well foregrounds
can be mitigated (see \citealt{Liu2014a}), so it is vital 
that we have a strong understanding of the statistics
of 21-cm fluctuations, even in light 
of next-generation instruments.
There are also a large number of astrophysical parameters 
(many of which are degenerate with each other, 
see for example \citealt{Greig2015a} and \citealt{Pober2015})
for which we have no constraints on in the
high-redshift Universe. 
Therefore we must also fully 
investigate all possibilities for the range 
of physics we might observe.

\subsection{Overview of this work}
Much attention has been focussed on measuring
the 21-cm power spectrum, which
has been shown to be rich with information
(for example, 
\citealt{Furlanetto2004a, Zaldarriaga2004, 
Mellema2006, Lidz2008, Pritchard2008, Santos2008, Signal2010, Friedrich2011, Mesinger2013a, Sobacchi2014a, Greig2015a}). 
However, given the challenging nature of these observations 
it is also worth considering one-point statistics.
One-point statistics have been shown to
be information rich, are simpler to interpret,
and will be differently 
affected by foregrounds (e.g. 
\citealt{Furlanetto2004, Wyithe2007a, Harker2009, 
Ichikawa2010, Watkinson2014, Watkinson2015}).

The sensitivity of the 21-cm one-point statistics to 
coupling and heating has not yet been studied in detail.
In this paper we investigate
the sensitivity of these statistics 
to the X-ray efficiency and spectral index
using semi-numerical simulations.
In doing so, we lift the assumption that $T_{\rm s}\gg T_{\textsc{cmb}}$
(which is often made when simulating reionization)
to study the impact of different X-ray properties
on the 21-cm moments during the EoR.

We note that during the writing of this paper a 
similar work by \citet{Shimabukuro2014}
was submitted to MNRAS. 
Our work differs in that our simulated boxes
are bigger (theirs are 200\,Mpc, ours are 600\,Mpc).
This is of particular importance in studying X-rays
because of their long mean free path,
which can be up to hundreds of Mpc
(see \citealt{McQuinn2012}).
We also include peculiar velocities in our simulations. 
Our paper includes several additional elements:
\begin{packed_item}
\item We present detailed analysis of the
impact of X-ray processes on reionization. 
In particular, we show that the evolution
of the skewness is altered in the case of
late X-ray heating, providing a useful
feature for constraining such models.
\item The impact of 
different values for the spectral index is studied,
identifying a degeneracy between X-ray efficiency and spectral index
(because both alter the amplitude of the 21-cm variance).
This degeneracy may be broken through
observations of the 21-cm skewness.
This provides sensitivity to the X-ray spectral hardness.
\item Finally we consider the
prospects for constraining X-ray source
properties with current and future generations
of radio telescope. In particular, we establish that even if 
foreground removal is not possible,
the 21-cm variance
can be accurately measured
using foreground avoidance techniques. 
\end{packed_item}

The paper is structured as follows:
In Section \ref{sec:sims} we describe our simulations;
in Section \ref{sec:coup_heat} we study the evolution 
of the variance and skewness during the epochs
of WF coupling and X-ray heating;
we then look at the impact of X-ray 
processes on the moments during
reionization in Section \ref{sec:EoR};
in Section \ref{sec:obs} we consider the observational
prospects for constraining the nature
of WF coupling, X-ray heating, and reionization
using the moments; 
finally in Sections \ref{sec:Disc} and \ref{sec:Conc} we discuss caveats of
our work and conclude.

\section{Simulation overview}\label{sec:sims}

We use 
the latest public release version of
\cmfast (v1.12) for this work.
For details 
of this simulation we refer the reader to 
\citet{Mesinger2007} and \cite{Signal2010};
however for convenience we will summarise 
the main points. 

The code uses the Zel'dovich Approximation \citep{Zel'dovichYa.B.1970}
applied to linear-density fields
to generate evolved density ($\delta$) and velocity ($\d v_{\rm r}/\d r$)
fields. 
The excursion-set formalism of \citet{Furlanetto2004}
can then be applied to the evolved density fields to generate 
neutral-fraction ($x_{\textsc{hi}}$) fields to model UV-driven ionizations.
The offset of the brightness temperature\footnote{
In radio observations, the radiation intensity $I_\nu$
is described in terms of 
brightness temperature, $T_{\rm b}$, defined such that $I_\nu=B(T_{\rm b})$; 
$B(T)$ is the Planck black-body spectrum and is well approximated by the 
Rayleigh-Jeans formula at the frequencies 
relevant to reionization studies.}
relative to that of the CMB ($\delta T_{\rm b}$) 
can then be calculated (assuming $T_{\rm s}\gg T_{\textsc{cmb}}$) according to,

\begin{equation}
\begin{split}
\delta T_{\rm b}=&\frac{T_{\rm s}-T_{\textsc{cmb}}}{1+z}(1-e^{-\tau_{\nu_0}})\,,\\
\approx&\,27\,\frac{T_{\rm s}-T_{\textsc{cmb}}}{T_{\rm s}}\,x_{\textsc{hi}}(1+\delta)\left[\frac{H(z)/(1+z)}{\d v_{\rm r}/\d r}\right]\\
&\times \left(\frac{1+z}{10}\frac{0.15}{\Omega_{\rm m}h^2}\right)^{1/2}\left(\frac{\Omega_{\rm b}h^2}{0.023}\right) \rm mK
\,.\\ \label{eqn:brightTemp}
\end{split}
\end{equation} 

\noindent The cosmological parameters in this 
equation are the Hubble parameter $H(z)$,
the matter $\Omega_{\rm m}$, 
and baryon density parameters $\Omega_{\rm b}$
(where $\Omega_i=\rho_i/\rho_{\rm c}$ and
$\rho_{\rm c}$ is the critical density required for
flat universe). 

Throughout this paper we adopt a standard 
$\Lambda$CDM cosmology as constrained by Planck, 
i.e. $\sigma_8=0.829$,
$h=0.673$, $\Omega_{\rm m}=0.315$, $\Omega_{\Lambda}=0.685$, 
$\Omega_{\rm b}=0.049$ and $n_{\rm s}=0.96$ \citep{Ade2014}.
Our simulations produce 3D boxes with side $L=600\,$Mpc 
and cubic pixels of side $l_{\rm pix}=1\,$Mpc
(initial conditions are calculated at double this
resolution).
All lengths are co-moving unless otherwise stated.

\subsection{X-ray heating and ionizations} 
This paper focuses on the effects
of X-ray heating, as such it is essential
that the details of 
spin temperature fluctuations are included. 
\cmfast calculates the spin temperature 
according to \citet{Field1958, Hirata2006a}, 

\begin{equation}
\begin{split}
T_{\rm s}^{-1} &= \frac{T_{\textsc{cmb}}^{-1} + x_\alpha T^{-1}_\alpha +x_{\rm c}T^{-1}_{\rm K} }{1 + x_\alpha + x_{\rm c}}\,,\\
 & \approx \frac{T_{\textsc{cmb}}^{-1} + x_\alpha T^{-1}_\alpha}{1 + x_\alpha}\,.\\
\label{eqn:spinT}
\end{split}
\end{equation}

\noindent The WF-coupling 
coefficient is defined as $x_\alpha = 1.7\times 10^{11}(1+z)^{-1}S_\alpha J_\alpha$
pcm$^{-2}$s$^{-2}$Hz$^{-1}$sr$^{-1}$, where
$S_{\alpha}\sim 1$ is a factor correcting for detailed atomic physics
and $T_\alpha$ is the Lyman-$\alpha$ colour temperature.
The second equality of Equation \ref{eqn:spinT} follows because 
the collisional coupling coefficient 
$x_{\rm c}\approx 0$ during the epoch relevant to this work.
Both $S_{\alpha}$ and $T_\alpha$ are calculated 
according to \citet{Hirata2006a}. 
For the purpose of our analysis, we define an `effective'
coupling co-efficient as 
$x_{\mathrm{eff}} =x_\alpha/(1+x_\alpha)$ assuming $S_\alpha=1$.

The Lyman-$\alpha$ flux ($J_\alpha$) has two main origins, 
(1) X-ray excitation (via photo-electrons)
of neutral hydrogen and (2) direct stellar 
emission between the Lyman limit and the 
Lyman-$\alpha$ line. 
Lyman-$n$ photons are redshifted into
Lyman-$\alpha$ resonance;
therefore, $J_\alpha$ is
calculated by integrating the 
photon contribution (as produced via mechanisms 1 and 2) 
over a series of concentric spherical redshift shells
surrounding each pixel.
For more detail than provided here,
we refer the interested
reader to the \cmfast \,literature listed
at the beginning of this section as well as 
\citet{Hirata2006a, Pritchard2006}; and \citet{Pritchard2007a}.

Where WF coupling is saturated,  
$x_{\rm \alpha}\gg T_{\textsc{cmb}}$ 
and $T_{\rm s}^{-1}  \approx  T^{-1}_\alpha$;
in this regime, the kinetic temperature will
be tightly coupled to $T_\alpha$ 
due to the repeated scattering of the Lyman-$\alpha$ photons
by hydrogen atoms \citep{Field1958}.
The kinetic temperature (outside of H\thinspace\textsc{ii}
regions at position 
$\boldsymbol{x}$ and at redshift $z$) is calculated 
by tracking the heating history for that position.
This can be calculated using the evolution of 
$T_{\rm k}(\boldsymbol{x}, z')$
which is predicted by,

\begin{equation}
\begin{split}
\frac{\d T_{\rm k}(\boldsymbol{x}, z')}{\d z'} 
=& \frac{2}{3k_{\rm \textsc{b}}(1+x_e)}\frac{\d t}{\d z'}\sum_p\epsilon_p + \frac{2T_{\rm k}}{3n_{\rm b}}\frac{\d n_{\rm b}}{\d z'} \\
&- \frac{T_{\rm k}}{1 + x_e}\frac{\d x_e}{\d z'}\,,\\
\label{eqn:Tk}
\end{split}
\end{equation}
where
\begin{equation}
\begin{split}
\frac{\d x_{e}(\boldsymbol{x}, z')}{\d z'} 
=& \frac{\d t}{\d z'}\left[ \Gamma_{\rm ion} - \alpha_{\rm A}Cx_{e}^2n_{\rm b}f_{\rm \textsc{h}}\right]\,.\\
\label{eqn:xe}
\end{split}
\end{equation}

\noindent In calculating the kinetic temperature,
\cmfast must also calculate the
local ionized fraction $x_{e}(\boldsymbol{x}, z')$,
which depends on the 
total baryon number density $n_{\rm b}$, 
the ionization rate per baryon $\Gamma_{\rm ion}$,
the case-A recombination co-efficient $\alpha_{\rm A}$,
the clumping factor 
$C=<n_{\rm \textsc{h}}^2>/<n_{\rm \textsc{h}}>^2$ 
(where $n_{\rm \textsc{h}}$ describes the hydrogen 
number density and we set $C=2$), 
 and
finally the hydrogen number fraction 
$f_{\rm \textsc{h}}$. 
In addition, the kinetic 
temperature depends on the 
heating rate ($\epsilon_p$) 
per baryon\footnote{The heating rate ($\epsilon_p$) has units erg s$^{-1}$.}
for process $p$ (for our discussion the
dominant process is X-ray heating)
and the Boltzmann constant $k_{\rm \textsc{b}}$.

To calculate X-ray heating and ionization rates at redshift $z'$
one must integrate over the full range of frequencies
for which photons can contribute energy to these processes. 
Furthermore, to account for redshifted photons, 
another integral over redshift 
($z''$)\footnote{Implicit is the assumption that $z''>z'$.} 
is required.
The X-ray luminosity of sources 
is assumed to be well described by a power law, 
i.e. $L_{\rm emitted} \propto (\nu/\nu_0)^{-\alpha}$ 
(where $\alpha$ is the spectral index discussed in Section \ref{sec:conts})
and $h \nu_0$ is the lowest energy X-ray
that can escape into the IGM). 
\cmfast assumes that the total X-ray 
emission rate per second ($\d \protect\dot{N}/\d z''$) from a 
spherical shell bounded by the redshift
interval $z''$ to $z''+\d z''$ is the product
of the number of
X-ray photons per solar mass in stars $\zeta_{\textsc{x}}$
(the X-ray efficiency parameter) and the star formation 
rate in that shell (SFR$_{z''}$) 
(i.e. $\d \protect\dot{N}/\d z'' = \zeta_{\textsc{x}} \mathrm{SFR}_{z''}$).\footnote{The star formation 
rate
SFR$_{z''}=\rho_{\rm crit, 0}\Omega_{\rm b}f_*[1+\overline{\delta}^{R''}_{\rm nl}(z'')] \d V/\d z''\d f_{\rm coll}(z'', R'')/\d t$, 
where $\d V(z'')$ is the co-moving volume element at $z''$, $\delta^{R''}_{\rm nl}(z'')$ is the PT-evolved density field 
smoothed on scale $R''$, $f_*$ is the fraction of baryons
converted to stars, $f_{\rm coll}$ is the collapsed fraction
(which is calculated as described in Section \ref{sec:ion}),
and $\rho_{\rm crit, 0}$ is the critical density at $z=0$.}
The arrival rate at position $\boldsymbol{x}$ and redshift $z'$
from sources between $z''$ and $z''+\d z''$
is then,
\begin{equation}
\begin{split}
  \frac{\d \phi(\boldsymbol{x}, \nu ,z',z'')}{\d z''} = \frac{\d \protect\dot{N}}{\d z''}\alpha\nu_0^{-1}\left(\frac{\nu}{\nu_0}\right)^{-\alpha-1}\left(\frac{1+z''}{1+z'}\right)^{-\alpha-1}\mathrm{e}^{-\tau_{\rm x}}\,,
\end{split}
\end{equation}
where $\tau_{\rm x}$ is the optical depth of X-rays. 
In calculating the mean free path of X-rays, fluctuations
of the IGM state are ignored.
Note that this is very inaccurate
during the advanced stages of reionization,
when there are large ionized regions in an otherwise neutral
IGM. With this approximation, the heating rate
due to X-rays at position $\boldsymbol{x}$ 
and redshift $z'$ is calculated as,
\begin{equation}
\begin{split}
\epsilon_{\rm X}(\boldsymbol{x}, z') =& \int_{\mathrm{Max[\nu_0,\nu_{\tau=1}]}}^\infty \d \nu \\
&\times \sum_i (h\nu-E_i^{\rm th})f_{\rm heat}f_ix_i\sigma_i \\
&\times \int_{z'}^\infty\d z'' \frac{\d \phi/\d z''}{4\pi r_{\rm p}^2}\,, 
\end{split}\label{eqn:from}
\end{equation}
\noindent where $r_{\rm p}$ is the proper (null geodesic) 
separation of $z'$ and $z''$. 
Under the same assumption, the ionization 
rate due to X-rays may be described by,
\begin{equation}
\begin{split}
\Gamma_{\rm ion}(\boldsymbol{x}, z) =& \int_{\mathrm{Max[\nu_0,\nu_{\tau=1}]}}^\infty \d \nu \sum_i f_ix_i\sigma_iF_i\\
&\times \int_{z'}^\infty \d z'' \frac{\d \phi/\d z''}{4\pi r_{\rm p}^2} \,,\\
\end{split}
\end{equation}
\begin{equation}
\begin{split}
F_i =&(h\nu-E_i^{\rm th})
\left(
\frac{f_{\rm ion, H\textsc{i}}}{E_{\rm H\textsc{i}}^{\rm th}} + 
\frac{f_{\rm ion, He\textsc{i}}}{E_{\rm He\textsc{i}}^{\rm th}} +
\frac{f_{\rm ion, He\textsc{ii}}}{E_{\rm He\textsc{ii}}^{\rm th}}
\right) + 1\,.\\
\end{split}\label{eqn:to}
\end{equation}

In Equations \ref{eqn:from} to \ref{eqn:to}
a sum is taken over
the species $i=$H\thinspace\textsc{i}, He\thinspace\textsc{i},
He\thinspace\textsc{ii}; $x_i$ is the ionization 
fraction for the species\footnote{$x_i = 1 - x_e$ for
H\thinspace\textsc{i} and He\thinspace\textsc{i}; $x_i = x_e$ 
for He\thinspace\textsc{ii}.}, $f_i$ is the species 
number fraction, $\sigma_i$ is the ionization
cross section, and $E_i^{\rm th}$ 
is the species ionization threshold energy.
The fraction of the primary electron's
energy that is transferred to heat 
and secondary ionizations is described 
by $f_{\rm heat,i}$ and $f_{\rm ion,i}$ respectively
for each species.
The unity term in $F_i$ accounts for
primary ionizations of species $i$.
The heating and ionization rate are simplified further
(to speed up the calculation)
by assuming that no photons with an 
optical depth $\tau \le 1$ are
absorbed by the IGM and all photons
with $\tau > 1$ are \citep{Mesinger2007, Signal2010}.
 
\subsection{UV Ionizations} \label{sec:ion}
Ionizations by UV photons 
are calculated independently from, and following 
the simulation of X-ray heating and WF coupling. 
The code smooths iteratively around each pixel
in the box 
from a maximum radius $R_{\rm max}$ 
\footnote{We set $R_{\rm max}=30$\,Mpc based on 
the ionizing UV photon mean free path at the 
redshifts of interest, see 
\citet{Storrie-Lombardi1994, MiraldaEscude2003, Choudhury2008}.}
down to pixel scales $R_{\rm pix}$. 
At each smoothing step, on scale $R$, 
the condition 
$f_{\rm coll}(\mathbf{x}, z, R)\ge\zeta_{\rm uv}^{-1}$
is evaluated, if met the central pixel is marked
as ionized; if this condition is never met 
then the pixel is partially ionized, accounting 
for both UV ionizations, calculated as 
$x^{\mathrm{uv}}_{\mathrm{ion}}=\zeta_{\rm uv} f_{\rm coll}(\bmath{r},z,R_{\rm pix})$,
and partial ionizations due to X-rays
(calculated using Equation \ref{eqn:xe}).

The collapsed fraction $f_{\rm coll}$ is calculated 
using the prediction of the extended Press-Schechter 
formalism \citep{Bond1991,LaceyCedric1993},
with the minimum mass corresponding to a virial
temperature of 10$^4$ K (necessary for cooling by atomic hydrogen
to be effective).
The collapse fraction is normalised
so that its mean agrees with
that predicted by the parametrically fit
mass function of \citet{Jenkins2001}. 

The ionizing efficiency of stars is defined as
 $\zeta_{\rm uv} = f_{\rm esc}f_*N_{\gamma/b}(1+n_{\rm rec})^{-1}$,
where $f_{\rm esc}$ is the fraction of ionizing photons that 
escape,  
$N_{\gamma/b}$ denotes the number of ionizing photons
produced per baryon in stars, and finally $n_{\rm rec}$ is the 
typical number of times that hydrogen will have recombined.
The fraction of baryons converted to stars $f_*$
also impacts upon the estimation of SFR used 
in the X-ray heating and ionization rates. 
We set $\zeta_{\rm uv} = 16$ for consistency with
our previous publications, and
so that the 50\% point of reionization falls in
the redshift range to which first generation
instruments are most sensitive, whilst agreeing
with observational constraints on the EoR.
Our reionization model is thus optimistic.

Because UV photons have a short 
mean free path, it is assumed that they will carve out large ionized
regions in a mostly neutral IGM; it is useful then to 
consider the volume filling factor of these ionized
H\thinspace\textsc{ii} regions $Q_{\rm \textsc{hii}}$. 
The average
ionized fraction of the box, 
taking into account the X-ray
ionizations discussed above, is 
$\overline{x}_{\mathrm{ion}}=Q_{\rm \textsc{hii}} + (1-Q_{\rm \textsc{hii}})x_e$.

\section{Results}
\begin{figure}
  \centering
  \includegraphics{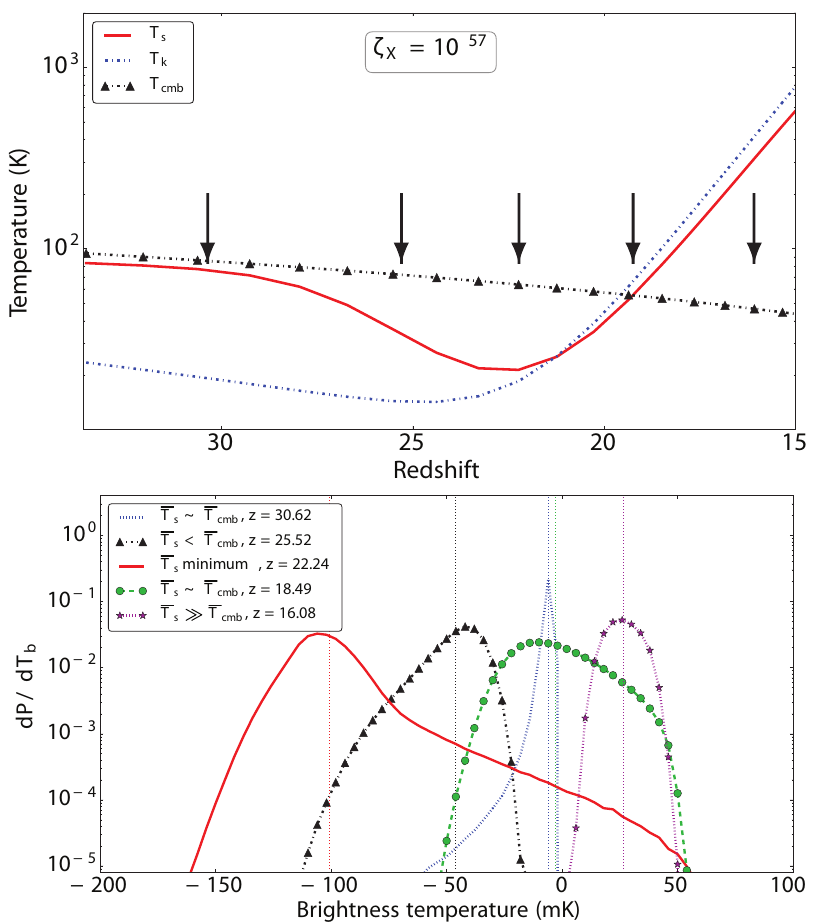}\\
  \caption{Redshift evolution of the spin  ($T_{\rm s}$), kinetic  ($T_{\rm k}$), and CMB ($T_{\textsc{cmb}}$) temperatures (top), and brightness-temperature PDF (bottom) at 
five key phases of its evolution for the fiducial model with $\zeta_{\textsc{x}}=10^{57}$ and $\alpha=1.5$.
All PDFs are from maps smoothed to 4\,Mpc.
Arrows on the top plot correspond to the redshifts
for which we present PDFs.
The nature and timing of these
features are model dependent;
in some extreme models it is possible that
not all of the five phases we describe will exist,
or that their PDFs will look quite different.
}
  \label{fig:TbPDF}
\end{figure}

In this work, we study the properties of X-ray sources
by varying the luminosity normalisation and spectral index.
The normalisation of the luminosity 
is parametrised through the X-ray
efficiency parameter $\zeta_{\textsc{x}}$ in \textsc{\small 21CMFAST }.  
We simulate $\zeta_{\textsc{x}} = [10^{55},10^{56},10^{57},10^{58}]$,
with $\zeta_{\textsc{x}} = 10^{57}$
(roughly 1.7 X-ray photons per
stellar baryon) as our fiducial model. 
This choice is consistent with low-redshift 
constraints on the total X-ray luminosity
per unit of star formation (with $f_* = 0.1$).
We then consider the hardness
of the X-ray background by varying the spectral index,
assuming values ranging from $\alpha=0.8$ 
(to approximate the spectrum produced by HMXB) to $\alpha=3.0$
(typical of a soft X-ray background as produced by the hot ISM).
We set $\alpha=1.5$ in our fiducial model,
as it is in the middle of the plausible range of
values for this parameter
and is representative of X-ray hardness
in the local Universe (see the discussion in Section \ref{sec:conts}).
Unless otherwise stated results are
from maps that have been smoothed and re-sampled to
produce pixels with side 4\,Mpc. This is to overcome
the impact of a discretisation effect (that occurs
through the creation of the non-linear density fields) on the moments
(see \citealt{Watkinson2014} for details).

Ignoring fluctuations in peculiar velocities
and at a fixed redshift (and cosmology), 
the drivers of brightness-temperature fluctuations
are the density, neutral-fraction 
($x_{\textsc{hi}} = 1 - x_{\mathrm{ion}}$) 
and spin-temperature fields; specifically,

\begin{equation}
\begin{split}
\delta T_{\rm b}\propto \mu \,x_{\textsc{hi}}\Delta \,,\\ \label{eqn:tbPropto}
\end{split}
\end{equation}

\noindent with $\mu = 1-T_{\textsc{cmb}}/T_{\rm s}$ and
$\Delta = 1+\delta$. At early times, before the
epoch of reionization, $x_{\textsc{hi}}\sim 1$
and fluctuations in $\mu$ and $\Delta$ dominate
the signal.
Therefore any evolution of the brightness temperature
away from that of the density field will be
due to correlations between 
$\mu$ and $\Delta$. 
As such we can gain insight by looking at the 
cross averages of these quantities,
which can be broken up as follows,

\begin{equation}
\begin{split}
\langle \mu \,x_{\textsc{hi}}\Delta\rangle &= \overline{x}_{\rm \hi}-  \left\langle \delta \frac{T_{\textsc{cmb}}}{T_{\rm s}}x_{\textsc{hi}}\right\rangle - \left\langle \frac{T_{\textsc{cmb}}}{T_{\rm s}}x_{\textsc{hi}}\right\rangle  + \left\langle \delta x_{\textsc{hi}}\right\rangle \,;\\ 
\mathrm{or\,when }\, x_{\textsc{hi}}= 1\mathrm{:}\\
\langle \Delta \, \mu \rangle &= 1 - \left\langle \delta \frac{T_{\textsc{cmb}}}{T_{\rm s}}\right\rangle - \left\langle \frac{T_{\textsc{cmb}}}{T_{\rm s}}\right\rangle  
\,.\\ 
\label{eqn:cross_aves}
\end{split}
\end{equation}

\subsection{Coupling and heating epochs}\label{sec:coup_heat} 

\begin{figure*}
\begin{minipage}{176mm}
\begin{tabular}{c}
  \begin{overpic}[trim=0.0cm 0.3cm 0cm 0.0cm, clip=true]{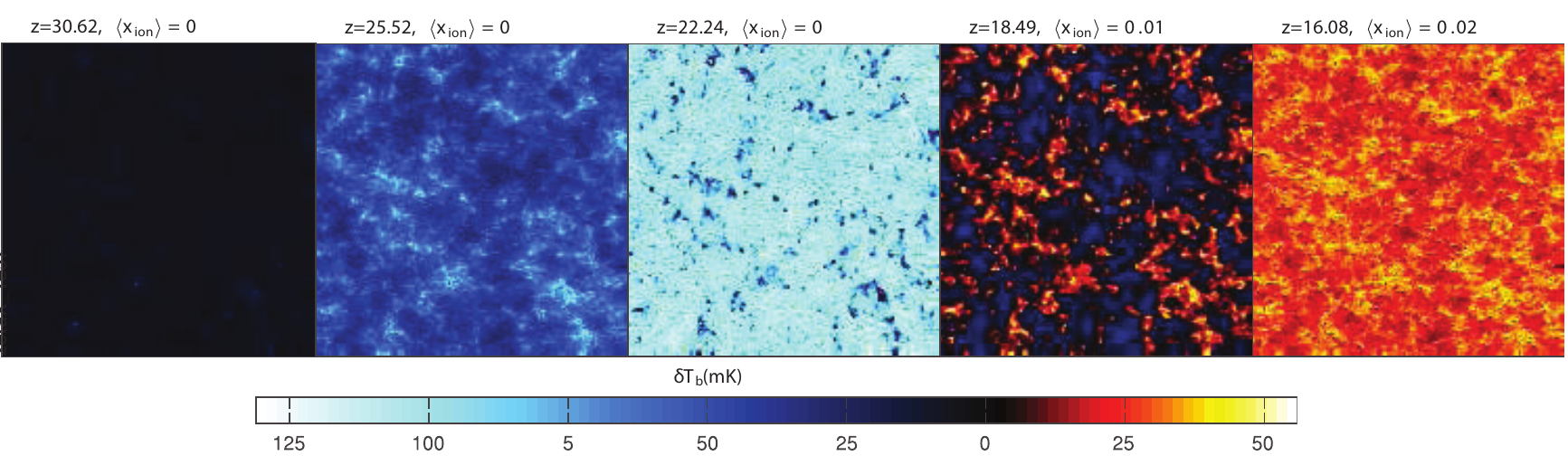}
  \put(83,-6){-125}
  \put(128,-6){-100}
  \put(175,-6){-75}
  \put(219,-6){-50}
  \put(264,-6){-25}
  \put(313,-6){0}
  \put(355,-6){25}
  \put(400,-6){50}
  \end{overpic}\\
\end{tabular}
\caption{Brightness-temperature maps of our fiducial
($\zeta_{\textsc{x}}=10^{57}$, $\alpha=1.5$) simulation.
Maps are from boxes of side $L= 600\,$Mpc smoothed 
to have pixels of (4 Mpc)$^3$ and presented with the pixel 
depth flattened into the page.
Maps correspond to the five key phases we describe in the text;
i.e. $z\sim 30$ where $\overline{T}_{\mathrm{s}}\sim \overline{T}_{\textsc{cmb}}$ (left), 
$z\sim 25$ where $\overline{T}_{\mathrm{s}} < \overline{T}_{\textsc{cmb}}$ (middle-left),
$z\sim 22$ where $\overline{T}_{\mathrm{s}}$ is at its minimum (middle),
$z\sim 18$ where $\overline{T}_{\mathrm{s}} \sim \overline{T}_{\textsc{cmb}}$ (middle-right) and 
$z\sim 16$ where $\overline{T}_{\mathrm{s}}\gg \overline{T}_{\textsc{cmb}}$ (right).
The amplitude of the signal is model dependent,
the more efficient the X-ray production the less
the amplitude in absorption. If the X-ray efficiency
is high enough, it is even possible
that the signal will never enter a phase of
absorption as seen in this fiducial model.
}\label{fig:dTSlice}
\end{minipage}
\end{figure*}

\begin{figure}
  \centering
  \includegraphics{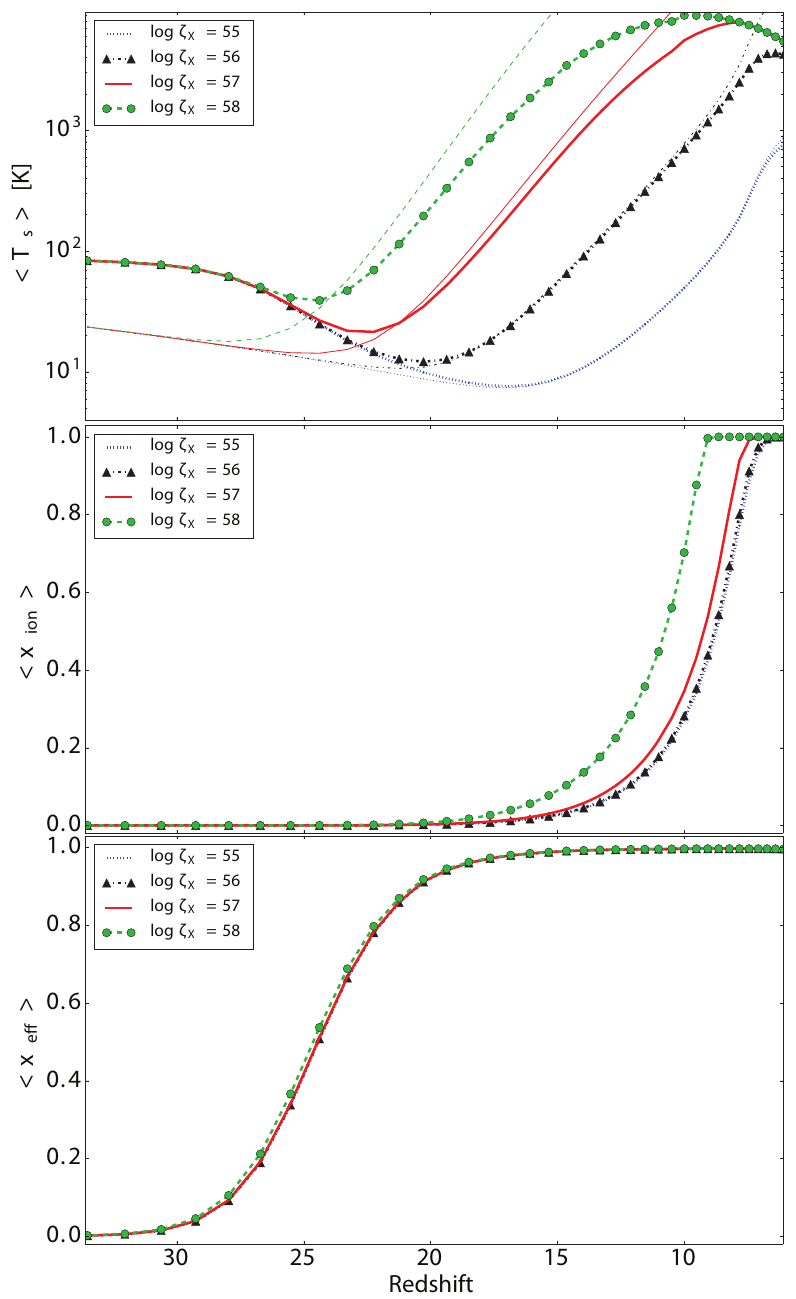}
  \caption{Redshift evolution of: spin (thick lines) and kinetic (thin lines) temperatures (top); ionized fraction (middle); and effective coupling coefficient (bottom).
$\alpha=1.5$ for all models.
Only in the extremely X-ray-efficient
$\zeta_{\textsc{x}}=10^{58}$ model do X-ray ionizations
contribute substantially to reionization.
All other models exhibit a very similar, UV-driven,
reionization history.
However, the evolution of the spin temperature
is different between all models.}
  \label{fig:gen_evo}
\end{figure}
  
Before considering the evolution of the 
21-cm moments, we can build some insight
by looking at the probability density function (PDF) of the brightness temperature.
The 
top plot of Fig. \ref{fig:TbPDF} 
shows the redshift evolution
of the three temperature components relevant to $\delta T_{\rm b}$:
the kinetic temperature ($T_{\rm k}$; blue-dashed line),
the CMB temperature ($T_{\textsc{cmb}}$; black-dashed line w/triangles)
and the spin temperature ($T_{\rm s}$; red solid line).
The bottom plot of the same figure 
shows the shape of the $\delta T_{\rm b}$ PDF at five important
phases of the brightness-temperature's evolution for our fiducial model
(see \citealt{Shimabukuro2014}
for discussion of the $1-T_{\textsc{cmb}}/T_{\rm s}$ PDF,
which agrees with the interpretation we present below).\footnote{We choose to plot
the PDFs with a log y-axis as we find it better for visualizing
skewness in the distributions.}

For reference, the associated brightness-temperature maps
are presented in 
Fig. \ref{fig:dTSlice} along with the redshift evolution
of spin temperature, ionized fraction and $x_{\mathrm{eff}}$
in Fig. \ref{fig:gen_evo} (top, middle and bottom respectively).

\begin{packed_item}
\item \textbf{$\boldsymbol{\overline{T}_{\rm s}\sim \overline{T}_{\textsc{cmb}}}$,
$\boldsymbol{z\sim 30}$}
(Blue dot-dashed line in Fig. \ref{fig:TbPDF} bottom): 
WF coupling begins almost
immediately in our simulations and is positively
correlated with the density field 
(this can be seen in the plot of
the cross average of the `effective' WF coupling coefficient
with density as a function of redshift in the
bottom of Fig. \ref{fig:variance}). 
As such the spin temperature in overdense
regions (near the sources of Lyman-$\alpha$ radiation)
is becoming coupled to the gas temperature 
(which is cooling adiabatically with the 
expansion of the universe).
As a result the mean spin 
temperature drops below
that of the CMB. 
This process
produces a negatively skewed 
brightness-temperature PDF 
which is quite sharply peaked 
with the weight of the distribution 
towards more negative
brightness temperatures.
At this point, brightness-temperature fluctuations are 
dominated by fluctuations in the density and Lyman-$\alpha$
flux. 

\item \textbf{$\boldsymbol{\overline{T}_{\rm s} < \overline{T}_{\textsc{cmb}}}$, 
$\boldsymbol{z \sim 25}$}
(black dashed line with triangles in Fig. \ref{fig:TbPDF} bottom):
The Lyman-$\alpha$ coupling coefficient and 
the density field are most strongly correlated around this
epoch for all models presented in this paper
(again refer to 
Fig. \ref{fig:variance} bottom). 
As the Lyman-$\alpha$ coupling becomes more effective
the spin temperature starts to evolve
more rapidly towards gas temperature, 
and the skewness of the PDF becomes 
less negative (as the statistics of the density
field become increasingly influential).
The variance is increasing during this phase. 

\item \textbf{$\boldsymbol{\overline{T}_{\rm s}}$ at its minimum, 
$\boldsymbol{z \sim 22}$}
(red solid line in Fig. \ref{fig:TbPDF} bottom):
Eventually the spin temperature reaches a minimum
just before coupling fully with 
the (now increasing) gas temperature.
From the PDF we can see that despite the 
average brightness temperature being at its minimum, 
some more extreme pixels are already 
in emission; i.e. coupling and X-ray
heating are both very strong in some pixels.
At this point, the PDF has a positive
skewness, primarily driven by fluctuations 
in the X-ray heating but amplified by
fluctuations in the WF-coupling.
This is because a region that is less
strongly coupled will have a spin temperature
closer to that of the CMB; 
a region that is both strongly coupled and more
heated than the mean will also result in a spin
temperature closer to that of the CMB.

\item \textbf{$\boldsymbol{\overline{T}_{\rm s} \sim \overline{T}_{\textsc{cmb}}}$ 
again, }$\boldsymbol{z \sim 18}$ (Green dashed line w/circles
in Fig. \ref{fig:TbPDF} bottom):
The spin temperature is now fully coupled
to the gas temperature and is thus increasing 
due to X-ray heating.
At this point, fluctuations in the X-ray heating are
dominating those of the brightness temperature.
The average brightness temperature is zero,
and fluctuations produce a relatively 
even distribution of pixels in emission and 
absorption; 
therefore the skewness is close to zero. The variance
is also decreasing as X-ray heating is becoming more
homogeneous.

\item \textbf{$\boldsymbol{\overline{T}_{\rm s} \gg \overline{T}_{\textsc{cmb}}}$, $\boldsymbol{z \sim 16}$} 
(Pink dotted line with stars in Fig. \ref{fig:TbPDF} bottom):
Eventually the spin temperature becomes much greater 
than the CMB temperature and heating fluctuations 
become unimportant. This results in a nearly
Gaussian distribution as the brightness-temperature
fluctuations are governed nearly entirely by those of the
density field. 
Reionization by UV photons is just becoming effective 
around this time. An earlier reionization model and/or
less efficient X-ray production could mean that this
Gaussian phase never occurs;
instead there may be a phase in which
fluctuations in both the heating
and ionization fields occur at the same time 
(as seen in the extreme $\zeta_{\textsc{x}} = 10^{55}$, 
which we describe at length in Section \ref{sec:EoR}).
\end{packed_item}

It is important to note that the PDFs 
described are
from our fiducial ($\zeta_{\textsc{x}}=10^{57}$, $\alpha=1.5$)
model. Thus, these five points may
be observed at different redshifts;
the evolution of the PDFs will also
vary quantitatively in different models.
Furthermore, if X-ray production is either extremely
efficient, or extremely inefficient, then the evolution
of the various temperatures and therefore the PDFs will
be qualitatively different from the fiducial model. 

\subsubsection{Efficiency of X-ray production}

\begin{figure}
  \centering
  \includegraphics[trim=0.15cm 0.0cm 0cm 0cm, clip=true]{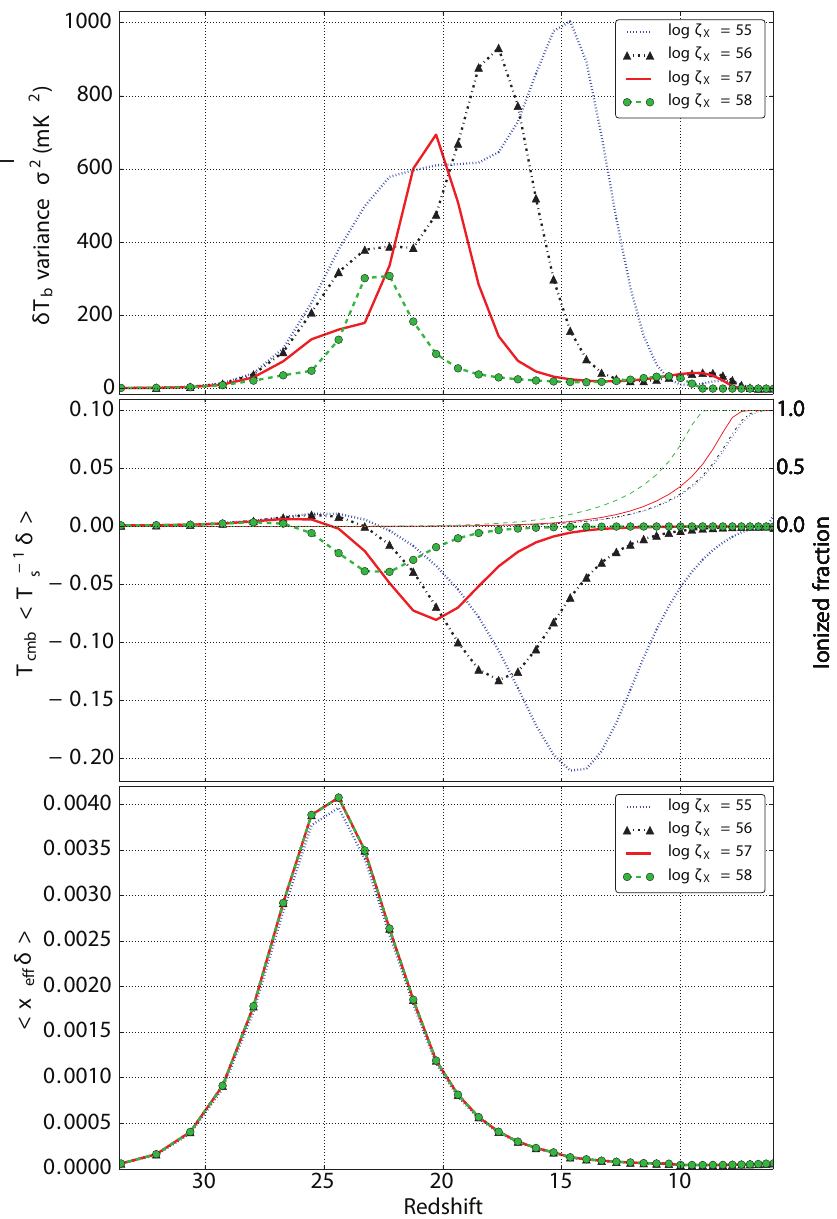}\\
  \caption{Redshift evolution of: brightness-temperature variance (top)
which exhibits two peaks prior to that of reionization; the first
is driven by $\langle x_{\mathrm{eff}} \delta\rangle$ (bottom) and the second
by $\langle (T_{\textsc{cmb}}/ T_{\rm s})\delta\rangle$ (middle).
Thin lines on the middle plot correspond to the evolution
of the ionized fraction for each model.
Statistics are calculated from maps resolved to 4\,Mpc,
and $\alpha=1.5$ for all models.}
  \label{fig:variance}
\end{figure}

Fig. \ref{fig:variance} (top)
shows the redshift evolution of the
brightness-temperature PDF's variance. 
The variance is zero at very high redshift
for all models.
It then increases with decreasing redshift,
driven by a slight positive correlation
between the density field and $T_{\rm s}^{-1}$;
i.e. the spin 
temperature is smaller in overdense regions, because WF-coupling
is strongest in the vicinity of sources and 
during this phase $T_{\rm k} < T_{\textsc{cmb}}$. 
This is illustrated by the evolution
of $\langle T_{\textsc{cmb}}\delta/T_{\rm s}\rangle$
shown in the middle plot of Fig. \ref{fig:variance}. 
The evolution of the variance plateaus 
briefly as the average spin temperature 
drops towards the average gas temperature (although note 
this is less evident in the $\zeta_{\textsc{x}}=10^{58}$ as X-ray
heating occurs so early).
Eventually an anti-correlation 
between the density field and $T_{\rm s}^{-1}$ develops.  
By this point, WF-coupling fluctuations are minimal 
(see the 
bottom plot of Fig. \ref{fig:variance})
and so this effect
is caused by the underdense regions being less heated 
by X-rays than those closer to sources;
i.e. the spin temperature is smallest in underdense regions
where there are less X-ray sources.
In all but the $\zeta_{\textsc{x}}=10^{55}$ model,
the variance is largest when this anti-correlation 
is maximized.
As we will see, the $\zeta_{\textsc{x}}=10^{55}$ model
enters this phase during the early stages
of the EoR, when
fluctuations in $x_{\textsc{hi}}$ are becoming influential.
However, even in this model the influence of $x_{\textsc{hi}}$
is small, so the amplitude and position of the 
variance's maximum should provide a constraint on 
the X-ray production efficiency. The extent of the
plateau that precedes it could provide insight into the relative
timing between the onset of WF-coupling and X-ray heating.

We can gain insight into the variance's strong dependence 
on the correlation between $T_{\rm s}^{-1}$ and $\delta$ by calculating
the variance of $\Delta \mu$. We find that,
\begin{equation}
\begin{split}
\sigma^2_{\Delta \mu} &= \left\langle \left( \delta \frac{T_{\textsc{cmb}}}{T_{\rm s}}
- \delta \right)^2 \right\rangle 
+ 2\left\langle \delta \left( \frac{T_{\textsc{cmb}}}{T_{\rm s}} \right)^2 \right\rangle \\
&- 2\left\langle \delta \frac{T_{\textsc{cmb}}}{T_{\rm s}} \right\rangle - \left\langle \delta \frac{T_{\textsc{cmb}}}{T_{\rm s}} \right\rangle^2 \\
&- 2\left\langle \delta \frac{T_{\textsc{cmb}}}{T_{\rm s}} \right\rangle \left\langle \frac{T_{\textsc{cmb}}}{T_{\rm s}} \right\rangle \,,\\
\end{split}\label{eqn:anal_vari}
\end{equation}
and see that the variance is 
only sensitive to $\langle T_{\textsc{cmb}}/T_{\rm s} \rangle$ 
in the final term where its influence will be suppressed 
by a factor of $\langle  T_{\textsc{cmb}} \delta/T_{\rm s} \rangle$. 

\begin{figure}
\centering
\includegraphics{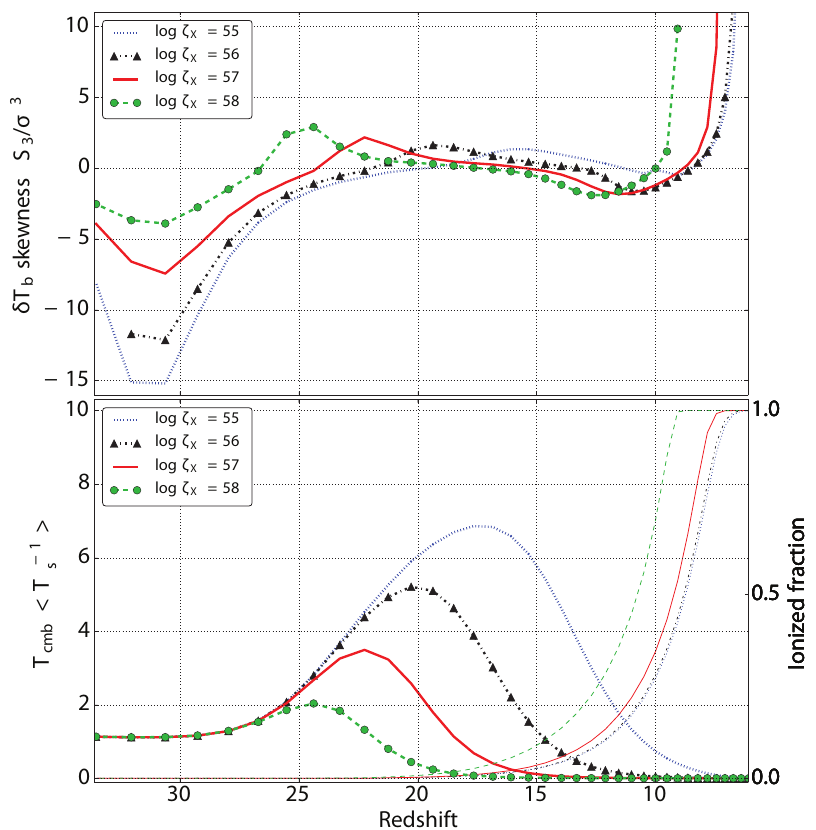}\\
  \caption{Redshift evolution of the brightness-temperature skewness (top)
which exhibits a minimum associated with the onset of WF coupling,
followed by a maximum that is driven by
$\langle T_{\textsc{cmb}}/T_{\rm s}\rangle$ 
(bottom).
Thin lines on the bottom plot correspond to the evolution of 
the ionized fraction for each model.
Statistics are calculated from maps resolved to 4\,Mpc,
and $\alpha=1.5$ for all models.
}
  \label{fig:skewness}
\end{figure}

\newcommand{\Trat}{{\frac{T_{\textsc{cmb}}}{T_{\mathrm{s}}}}}
\newcommand{\lang}{{\langle}}

In contrast, we find that the skew of $\Delta \mu$
(for which we include the full equation
in Appendix \ref{sec:appen})
is sensitive to both of these terms independently
and in combination. 
The position and amplitude of the
maximum in the skewness during 
this heating phase
is mainly sensitive to
$\langle T_{\textsc{cmb}}/T_{\rm s} \rangle$ 
as this factor dominates over 
$\langle T_{\textsc{cmb}}\delta/T_{\rm s} \rangle$.
This is clear from  Fig. \ref{fig:skewness}
where we plot
the skewness (top) and 
$\langle T_{\textsc{cmb}}/T_{\rm s} \rangle$ (bottom) as functions of redshift.
Initially the skewness becomes increasingly negative during
the early stages of WF-coupling. 
There is a universal minimum to the skewness of our models
at $z\sim31$ driven 
by fluctuations in the WF-coupling (the details of
which are unchanged between models)
drawing the spin temperature towards the 
lower kinetic temperature (see the discussion 
surrounding Fig. \ref{fig:TbPDF}).
The skewness increases from this minimum, 
becoming positive and reaching a maximum
as the average spin temperature 
(depicted in
the top plot of
Fig. \ref{fig:gen_evo})
reaches its lowest point. 
At this point, the $\mu$ parameter will be greatest and
so fluctuations in the spin temperature dominate. 

As previously discussed,
we see from the plot of
$\langle T_{\textsc{cmb}}/T_{\rm s} \rangle$ 
in the bottom plot of 
Fig. \ref{fig:skewness}
that
the amplitude of the X-ray heating skewness 
maximum is inversely proportional to that of 
$\langle T_{\textsc{cmb}}/T_{\rm s} \rangle$.
We find this to be due to contributions from negative 
$\langle T_{\textsc{cmb}}/T_{\rm s} \rangle^3$ terms becoming
more dominant as the spin temperature decreases (see Appendix \ref{sec:appen}). 

Note that in the $\zeta_{\textsc{x}} \le 10^{56}$ models,
the ionization field is 
becoming influential as the skewness reaches its
global maximum.
If we plot the redshift evolution of 
$\langle T_{\textsc{cmb}}x_{\textsc{hi}}/T_{\rm s}\rangle$
then we find a perfect correlation between the peak 
in skewness and the minimum of this cross average.
Even in such models, 
the high redshift maximum of the 
brightness-temperature skewness
should provide constraints on the point at which the 
spin temperature is minimum, and thus the 
efficiency of X-ray production.

\citet{Shimabukuro2014} show the brightness-temperature
variance and skewness for their fiducial
model ($\zeta_{\textsc{x}}=10^{56}$).
We mostly agree with their findings;
however, their plot of the brightness-temperature
variance only exhibits the X-ray heating peak
(note their plot does not show the redshifts associated
with reionization).
The peak we associate with WF coupling and the
plateau connecting it to the X-ray heating peak is
totally absent.
This may be because
their boxes
are small compared with ours.
However, it is most likely that
this difference is because \citet{Shimabukuro2014}
do not smooth their brightness-temperature
maps prior to measuring one-point statistics,
while we do.\footnote{There is
a discretisation effect in \textsc{\small 21CMFAST},
associated with the generation of the non-linear
density field, that must be smoothed out in
order to get a clean measure of the
brightness-temperature statistics \citep{Watkinson2014}.
This does not impact spin-temperature
simulations, which are the focus of \citet{Shimabukuro2014}.}.

\subsubsection{Hardness of the X-ray SED}
\begin{figure}
  \centering
  \includegraphics[trim=0.09cm 0.01cm 0.01cm 0.01cm, clip=true]{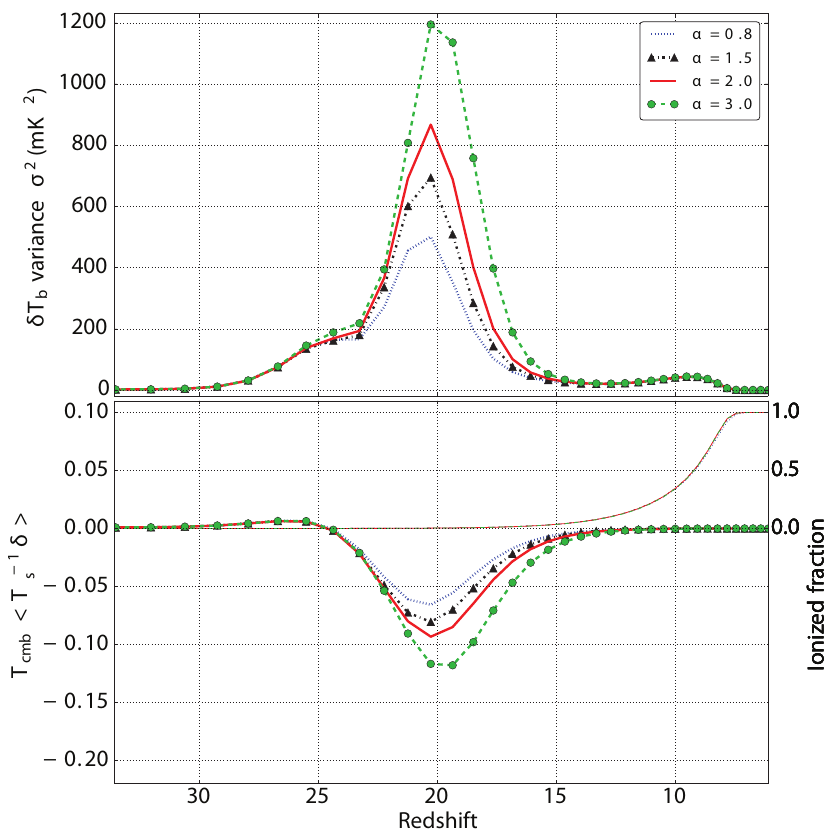}\\
  \caption{Evolution of variance with redshift (top) for various values
of the spectral index of the X-ray SED; $\zeta_{\textsc{x}}=10^{57}$ for all.
We find that suppression of the
$\langle \delta(T_{\textsc{cmb}}/ T_{\rm s})\rangle$
amplitude caused by a harder X-ray spectrum,
and seen in the bottom plot (for which the model key of the top plot applies),
reduces the variance.
Thin lines on the bottom plot correspond to the evolution of the ionized fraction for each model.
Statistics are calculated from maps resolved to 4\,Mpc.}
  \label{fig:TbVar_alpha}
\end{figure}

Fig. \ref{fig:TbVar_alpha} (top) shows
the redshift evolution 
of the brightness-temperature variance for
different choices of spectral index,
with $\zeta_{\textsc{x}}=10^{57}$.
The variance for the $\alpha=3.0$ (soft)
model is more than double that
of the $\alpha=0.8$ (hard) model. 
The softer the X-ray spectrum the greater the 
anti-correlation between the density field
and $T_{\rm s}^{-1}$ (i.e. the spin temperature 
is smallest in underdense regions).
This is evident in the bottom of
Fig. \ref{fig:TbVar_alpha}
where we plot the redshift evolution of
$\langle T_{\textsc{cmb}}\delta/T_{\rm s}\rangle$.
This is to be expected as soft X-rays
have a shorter mean free path than hard X-rays. 

\begin{figure}
  \centering
  \includegraphics{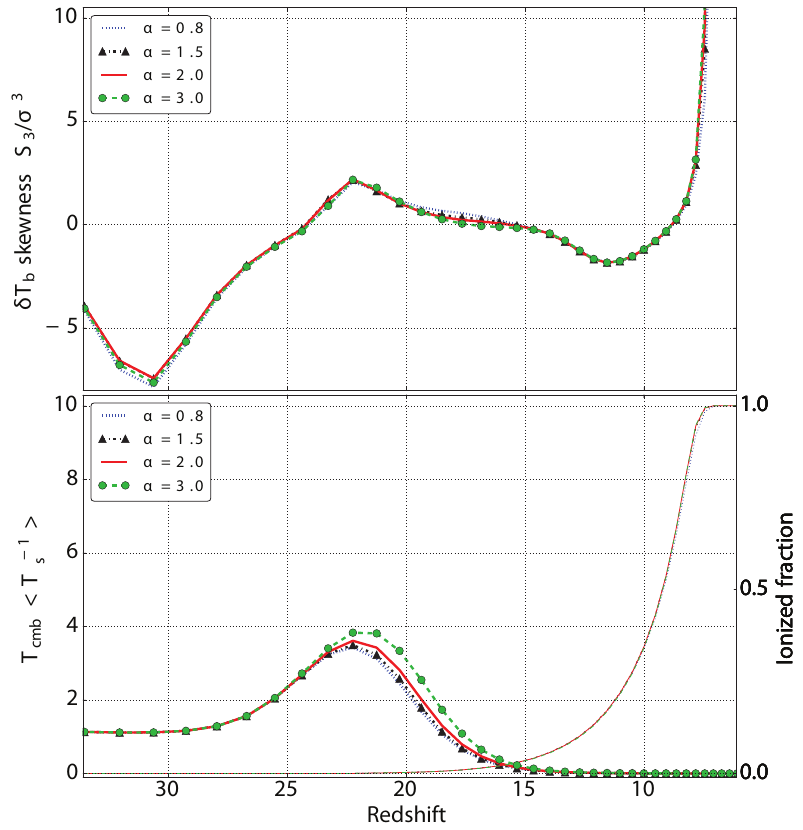}\\
  \caption{Evolution of the brightness-temperature skewness with
redshift (top) for various values of the spectral index of the assumed
X-ray SED; $\zeta_{\textsc{x}}=10^{57}$ for all.
As $\langle T_{\textsc{cmb}}/T_{\rm s}^{-1}\rangle$
(shown in the bottom plot as a function of redshift)
is insensitive to the hardness of the X-ray spectrum,
the skewness is not sensitive to the X-ray spectral index.
Thin lines on the bottom plot correspond to the evolution of the ionized fraction for each model.
Statistics are calculated from maps resolved to 4\,Mpc.
}
  \label{fig:TbSkew_alpha}
\end{figure}

The sensitivity of the variance amplitude to the spectral index
is degenerate with changes in amplitude
produced by different X-ray efficiencies. 
This degeneracy maybe broken as
the location and amplitude of the
skewness' X-ray heating peak is insensitive to
variations of the spectral index (as seen in
Fig. \ref{fig:TbSkew_alpha}
in which we plot the skewness for different spectral indices,
with $\zeta_{\textsc{x}}=10^{57}$).
This is because, the redshift at which
the spin temperature minimizes, and the difference 
between it and $T_{\textsc{cmb}}$, is driven
primarily by the efficiency of X-ray production.

We expect the insensitivity
of the skewness to the X-ray spectral hardness
to be relatively model independent across
the models we consider,
as $\langle T_{\textsc{cmb}}\delta /T_{\rm s}\rangle \ll \langle T_{\textsc{cmb}}/T_{\rm s}\rangle$ for all (see the bottom of Fig. \ref{fig:variance}
and Fig. \ref{fig:skewness}).
However, should the X-ray production be so efficient
that $\langle T_{\textsc{cmb}}/T_{\rm s}\rangle$ remains very small
during this phase, then the skewness would be
sensitive to $\langle T_{\textsc{cmb}}\delta /T_{\rm s}\rangle$,
and therefore the X-ray spectral hardness.
We conclude that if the efficiency can be constrained
using the skewness, then the amplitude of the variance
has potential for constraining the
spectral index of the X-ray SED.

\citet{Pacucci2014a} find the peak
amplitude of the large-scale ($k\sim 0.2\,$Mpc$^{-1}$) power spectrum
to be sensitive to the X-ray SED's spectral index,
but not the
efficiency of X-ray production. 
We do not recover this behaviour by
measuring the variance from maps smoothed on large scales.
We find instead that, for smoothing scales of order 60\,Mpc,
sensitivity to the spectral
hardness is lost whilst the amplitude 
remains sensitive to the efficiency of X-ray production.
This suggests that the variance and power spectrum
may be complementary in that the large-scale power spectrum 
can inform us on the spectral index
and the variance smoothed on large scales 
can provide constraints on the efficiency
of X-ray production.

\subsection{Epoch of reionization}\label{sec:EoR}

\begin{figure}
  \centering
  \includegraphics{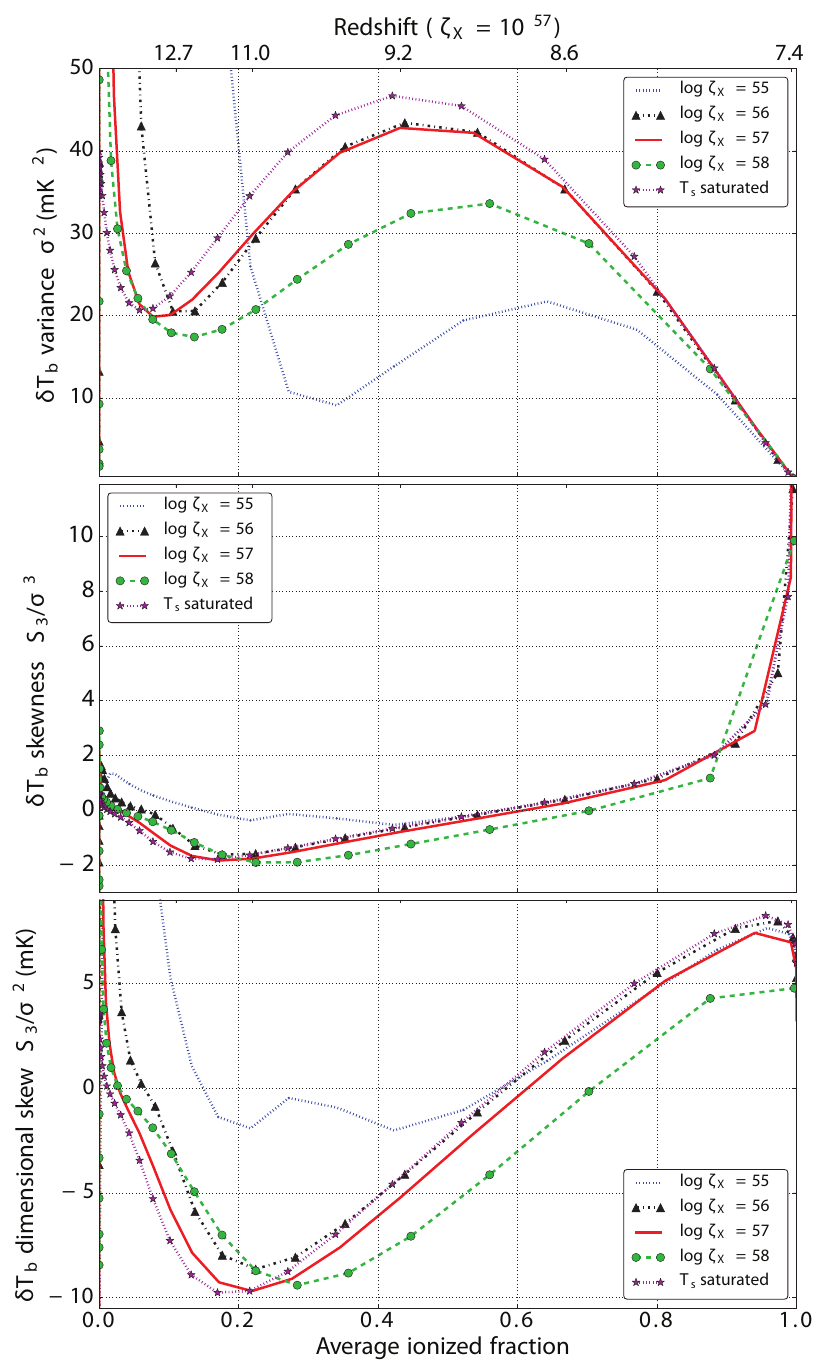}\\
  \caption{Brightness-temperature variance (top), skewness (middle) and dimensional skewness (bottom) as a function of ionized fraction to highlight features relevant to reionization.
X-ray ionizations are seen to reduce the amplitude of the variance
(see the green dashed line with circles)
by reducing the anti-correlation between density and neutral-fraction fields.
If heating is very late (see the blue-dashed line)
the turnover in the variance is no longer at the
mid-point of reionization, which could lead to
misinterpretation of the timing of the EoR.
The skewness is qualitatively different in such a model,
exhibiting a local maxima during the early
stages of reionization (note that the nature of this
feature will be extremely model dependent in this regime);
such a feature could be used to constrain late X-ray heating models.
Statistics are calculated from maps resolved to 4\,Mpc,
and $\alpha=1.5$ for all models.}
  \label{fig:TbReioMoments}
\end{figure}

\begin{figure}
  \centering
  \includegraphics{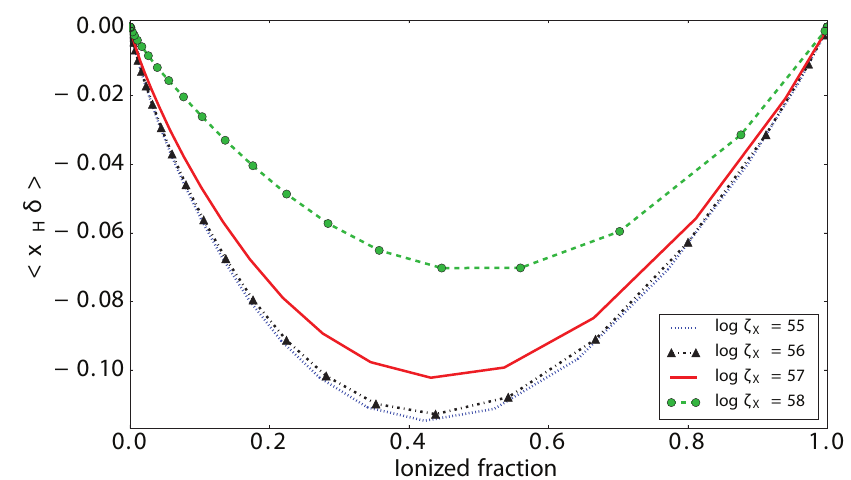}\\
  \caption{Evolution of the $\langle x_{\textsc{hi}} \delta \rangle$ cross average
as a function of ionized fraction; this confirms that it is
the decrease in the negativity of $\langle \delta x_{\textsc{hi}}\rangle$
that reduces the variance in the $\zeta_{\textsc{x}}=10^{58}$ model.
Statistics are calculated from maps resolved to 4\,Mpc,
and $\alpha=1.5$ for all models.}
  \label{fig:TbPDF_reio}
\end{figure}

When simulating reionization, it is often assumed that
the spin temperature is totally saturated and therefore
its fluctuations can be ignored.
We see in
Fig. \ref{fig:TbReioMoments} (in which
we plot the brightness-temperature moments as a function 
of ionized fraction) that this may
not be an appropriate assumption.
Note that \citet{Mesinger2013a} discuss
trends in the power spectrum's evolution at
$k=0.1\,$Mpc$^{-1}$ similar to those seen in the variance. 

\subsubsection{The impact of X-ray ionizations during reionization}
The variance
(Fig. \ref{fig:TbReioMoments} top)
for all our models 
is suppressed relative to
that of a simulation that uses identical initial conditions
but ignores spin temperature fluctuations
(labelled here as `$T_{\rm s}$ saturated').
This is due to partial ionization of neutral regions by X-rays.
X-ray ionizations are effective in both over and under-dense
regions, reducing the anti-correlation
between the density and neutral-fraction fields.
This is seen in Fig. \ref{fig:TbPDF_reio},
in which we plot $\langle x_{\textsc{hi}}\delta\rangle$ as a function of
ionized fraction, as the
anti-correlation reduces with
increasing X-ray efficiency.
Partial ionizations
also shift the mid-point maximum\footnote{The mid-point maximum refers
to a maximum in the evolution of the variance
during reionization.
This occurs as the average ionized fraction
of the Universe reaches 50\% when the spin temperature
is assumed to be saturated.}
in the variance 
to higher ionized fractions, as reionization 
is more advanced when driven just by UV radiation.
Such a shift is also seen in the minimum of the skewness and
dimensional skewness\footnote{the dimensional skewness 
refers to the skew normalised with $\sigma^2$ rather 
than $\sigma^3$, this was found by \citealt{Watkinson2014}
to be a more natural choice during reionization.}
associated with $\bar{x}_{\rm ion}\sim 0.2$
(see the middle and bottom plots of
Fig. \ref{fig:TbReioMoments} respectively).

The late-time features
of both skewness statistics at $\bar{x}_{\rm ion}\sim 0.95$,
i.e. the rapid increase in the skewness
as reionization advances,
and a turnover in the dimensional skewness,
are far more robust.
Although, for the highest efficiency we consider
($\zeta_{\textsc{x}}=10^{58}$) the late-time turnover
in the dimensional
skewness doesn't occur, as X-ray ionizations complete
reionization early relative to a UV-only model.
In the middle plot of Fig. \ref{fig:gen_evo},
we see that reionization completes at $z\sim 9$ in the $\zeta_{\textsc{x}}=10^{58}$
model; however, models which are mostly driven by UV ionizations
don't reach $\bar{x}_{\rm ion}\sim 0.95$ until $z\sim 7.5$.

\begin{figure}
  \centering
  \includegraphics{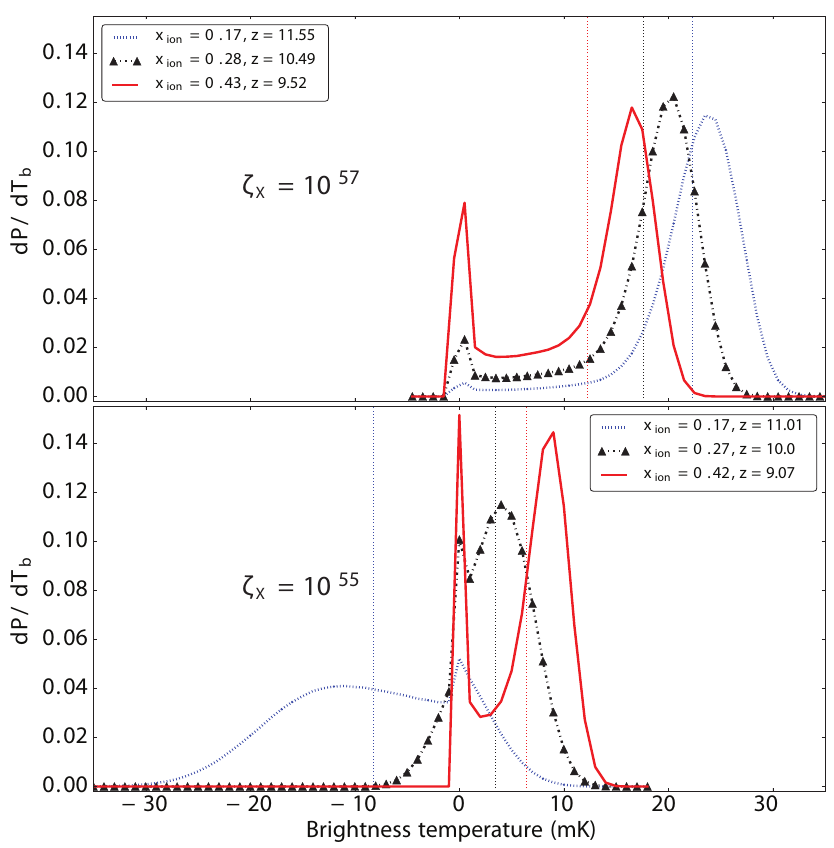}\\
  \caption{
Brightness-temperature PDF for redshifts 
(corresponding to $\bar{x}_{\mathrm{ion}}=$[0.17,0.28,0.43]) associated
with the local maximum feature seen in the skewness 
of the $\zeta_{\textsc{x}}=10^{55}$
model
for: $\zeta_{\textsc{x}}=10^{57}$ in the top plot (where the zero peak corresponding to
ionized pixels is separate from, and to the left of, the contribution from neutral regions);
and $\zeta_{\textsc{x}}=10^{55}$ in the bottom plot
(where the contribution from
neutral regions is merged with that of the zero peak).
Statistics are calculated from maps resolved to 4\,Mpc,
and $\alpha=1.5$ for all models.
}\label{fig:PDF_reio_extreme}
\end{figure}

\subsubsection{The impact of heating on interpreting signatures of reionization}
Following the X-ray heating-dominated phase, discussed in Section \ref{sec:coup_heat}, 
we see a rapid drop from the X-ray heating peak 
at low ionized fractions
(in agreement with the findings of \citealt{Mesinger2013a}).
If X-ray heating occurs relatively late as
in the $\zeta_{\textsc{x}} = 10^{55}$ model, the impact is 
dramatic as the drop from the heating peak
occurs during UV-driven reionization 
(occurring at $\bar{x}_{\rm ion}<0.3$).\footnote{\citet{Mesinger2013}
find that if X-ray heating is late enough, 
the heating and
reionization peaks can be merged into a single peak.}
Fig. \ref{fig:PDF_reio_extreme} shows the PDFs during this phase.
Unlike the $\zeta_{\textsc{x}} = 10^{57}$
model (where there is a clear distinction between
a positive brightness-temperature distribution and
a sharp spike at $\delta T_{\rm b}=0$), 
the brightness-temperature
distribution of the $\zeta_{\textsc{x}} = 10^{55}$ PDF 
extends to negative temperatures.
As a result, the contributions of neutral and ionized
regions to the PDF are no longer distinct in
brightness temperature. 
This reduces the variance and alters 
the skewness evolution, which exhibits a
local maxima as the skewness tends to zero when
$\overline{\delta T}_{\rm b}\sim 0$ in neutral regions.

Such signatures provide an 
opportunity to constrain the nature 
and timing of X-ray heating.
However they also complicate
interpretation of the variance and skewness
during reionization,
impacting our ability to 
constrain reionization using these moments.
For example, \citet{Patil2014} fit 
a function with a single peak to the variance of mock data,
in order to constrain parameters of reionization. 
Such an approach would
return misleading constraints, especially if late X-ray heating 
occurred. 

\citet{Ghara2015} note
this fact and suggest to use either a three peak
model (to model reionization, heating and coupling
peaks) or a redshift cut-off.
A redshift cut-off requires either prior
knowledge on the timing of heating and reionization
and/or throwing away information.
The data itself could be used to 
provide a prior on where a redshift cut should be made
(for example, model selection could be used to infer whether a
three, two or one peak model best describes the data in hand and
where the transitions from one to the next occur).
However, this would still be misleading
in the $\zeta_{\textsc{x}}=10^{55}$ model, as the drop from the heating
peak occurs over redshifts for which the
ionized fraction is $\sim 0.25$ and the peak
(usually associated with the mid point of reionization)
is at ionized fractions of between $0.6$ and $0.7$. 
We therefore conclude that it would be
prudent to use a parameter estimation
approach that uses simulations
to capture such subtleties. 
Unfortunately, this is particularly challenging as
simulations that include spin-temperature
fluctuations are computationally expensive.
Similar considerations would be necessary in 
constructing models of the skewness along the
lines of \citet{Patil2014}. 

These arguments are also relevant to MCMC parameter
estimation using simulations that assume the spin temperature
is saturated, such as that of \citet{Greig2015a} (who consider the
power spectrum rather than the variance). 
It would be interesting to test the code they 
describe (21\textsc{cmmc}) against
a mock dataset generated from models similar to
those we describe here to quantify the potential
bias we would suffer from ignoring the spin temperature
in performing parameter estimation.

\section{Observational prospects}\label{sec:obs}
To consider the effect of instrumental noise 
and foregrounds, we make use of the publicly 
available code 21\textsc{cmsense}\footnote{\href{https://github.com/jpober/21cmSense}{\protect\nolinkurl{https://github.com/jpober/21cmSense}}}
\citep{Pober2013, Pober2013a}.
We refer the readers to the 21\textsc{cmsense} literature
for details, but we will describe the main
points for completeness.

There are two main contributions to the 
error on the power spectrum: thermal noise
and sample variance. At lower redshifts shot
noise of the distribution of \hi must also 
be considered, but this term is neglected in this
analysis as it is found to be a sub-dominant effect,
even after reionization (see \citealt{Pober2013a}).
\citet{Pober2013a, Pober2013}
calculate the noise for the $k$-mode
measured by each individual baseline.
As such, for a given redshift, 
the power spectrum may be calculated by application 
of an inverse-variance-weighted summation,
for which the optimal estimator 
of the total noise error is,
\begin{equation}
\begin{split}
\delta\Delta^2(k) = 
\left\{  
\sum_i\frac{1}{[ \Delta^2_{\mathrm{\textsc{n}},i}(k) 
+  \Delta^2_{\mathrm{\textsc{sv}},i}(k) ]^2}
\right\}^{-\frac{1}{2}}\,.\\
\end{split}\label{eqn:noise_ivw}
\end{equation}
The thermal noise contribution for a $k$-mode labelled
by $i$ is given by
$\Delta^2_{\mathrm{\textsc{n}},i}(k)=X^2_iY_i[k_i^3/(2\pi^2)][\Omega_i/(2t_i)]T^2_{\mathrm{sys}}$.
In this expression $X^2Y\,[h^{-1}\,$Mpc$]$ translates observed units into cosmological distances;
$\Omega_i [\mathrm{steradians}]$ is the solid angle of the primary beam for a given baseline;
$T_{\mathrm{sys}}[\mathrm{K}]$ is the system temperature, a combination
of the sky and receiver temperatures,
(i.e. $T_{\mathrm{sys}} = T_{\mathrm{rec}} + T_{\mathrm{sky}}$);
and $t_i [\mathrm{hours}]$ is the integration time
for a given mode. 
The effect of the Earth's rotation
(relevant to a drift-scan observation mode\footnote{Note that SKA and LOFAR
can also perform tracked scans.})
is taken into account
when calculating the noise on an individual mode; 
i.e. different baselines may observe
the same mode at different times which increases
the integration time and therefore reduces thermal noise
(for a similar reason, redundant baselines are useful).
The sample variance error $\Delta^2_{\mathrm{\textsc{sv}},i}(k)$
is equivalent to the 21-cm power spectrum for that mode at a given redshift,
i.e. $\Delta^2_{\mathrm{\textsc{sv}},i}(k)  = \Delta^2_{21,i}(k) = k^3/(2\pi^2)P_{21}(k)$
where $P_{21}(k)$ is the 21-cm brightness-temperature
power spectrum.

As well as needing to beat down this error term,
there is also the issue of foregrounds which swamp the 
signal by several orders of magnitude.
By considering the Fourier transform along the frequency axis
of each mode independently (effectively measuring the
delay in signal arrival time between the two interferometer
elements that make up a baseline), 
\citet{Parsons2012} find that the spectrally smooth
nature of foregrounds mean that their contribution
will be confined to the region of delay space containing 
the maximum delays for a given baseline
(confining them to be below an `horizon limit'). On the other hand,
the 21-cm signal should exhibit unsmooth spectral characteristics
so that some contribution from the cosmological signal 
will be observed with smaller delays (i.e. above the `horizon limit').
This motivates the definition of $k_{\parallel,\mathrm{min}}$ and $k_{\perp,\mathrm{min}}$ 
below which foregrounds will dominate.
Because of the frequency dependence of interferometer
baselines, this `horizon limit' drifts to increasing values
of $k_{\parallel}$ with baseline length 
(i.e. with increasing $k_{\perp}$)
producing a `wedge' of foreground 
contamination in $k_{\parallel,\mathrm{min}}$-$k_{\perp,\mathrm{min}}$ 
parameter space.
Mathematically the $k_{\parallel,\mathrm{min}}$ `horizon limit'
may be described as \citep{Parsons2012},
\begin{equation}
\begin{split}
k_{\parallel,\mathrm{hor}} = \left(\frac{1}{\nu}\frac{X}{Y}\right)k_\perp
\,,\\
\end{split}\label{eqn:hor_limit}
\end{equation}
where $X$ and $Y$ convert angle and frequency
to co-moving distance respectively.

There are two main approaches to dealing
with the problem of foregrounds. 
One approach is to exploit the confinement
of foregrounds to the `horizon limits' described 
above and essentially ignore modes that fall 
outside of EoR window
(the region of $k_{\parallel}$-$k_{\perp}$  
space bounded by the `horizon limits');
see \citealt{Datta2010a, Vedantham2012, Morales2012, 
Thyagarajan2013, Hazelton2013, Liu2014}. 
When performing an
inverse-variance-weighted (IVW) summation over $k$-modes,
this is equivalent to assigning infinite noise
to modes that fall outside the EoR window.
In parallel, there is a great deal of effort going into 
actively removing foregrounds from observations;
these exploit the smooth spectral characteristics
of foregrounds to identify and  remove their contribution
(see \citealt{Wang2006a, Liu2011a, Paciga2011, Petrovic2011b, Chapman2012, Cho2012, Shaw2014}).

\begin{figure*}
\begin{minipage}{176mm}
\begin{tabular}{c}
  \includegraphics{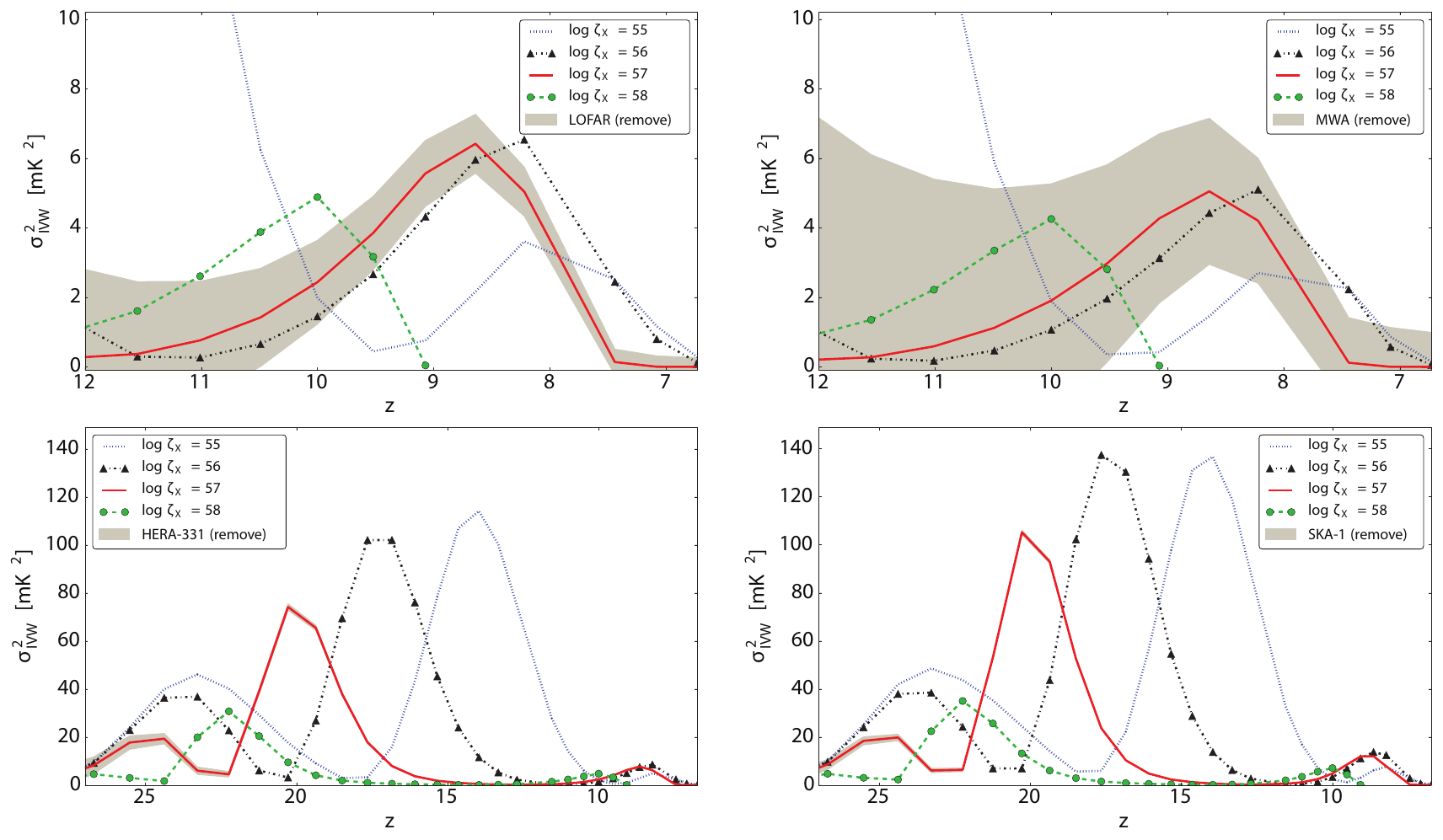}\\
\end{tabular}
\caption{Inverse-variance weighted variance as measured when foregrounds are removed. LOFAR (Top-left); MWA (top-right); HERA (bottom-left); SKA (bottom-right). In the interest of brevity we do not explicitly consider PAPER's performance, but PAPER-128 should be comparable in sensitivity to MWA for measuring the inverse-variance weighted variance.
In this best case scenario all instruments are seen to be capable of constraining reionization;
HERA and SKA would also tightly constrain WF-coupling and X-ray heating.
}
\label{fig:simCardVariance}
\end{minipage}
\end{figure*}

\begin{figure}
  \centering
  \includegraphics{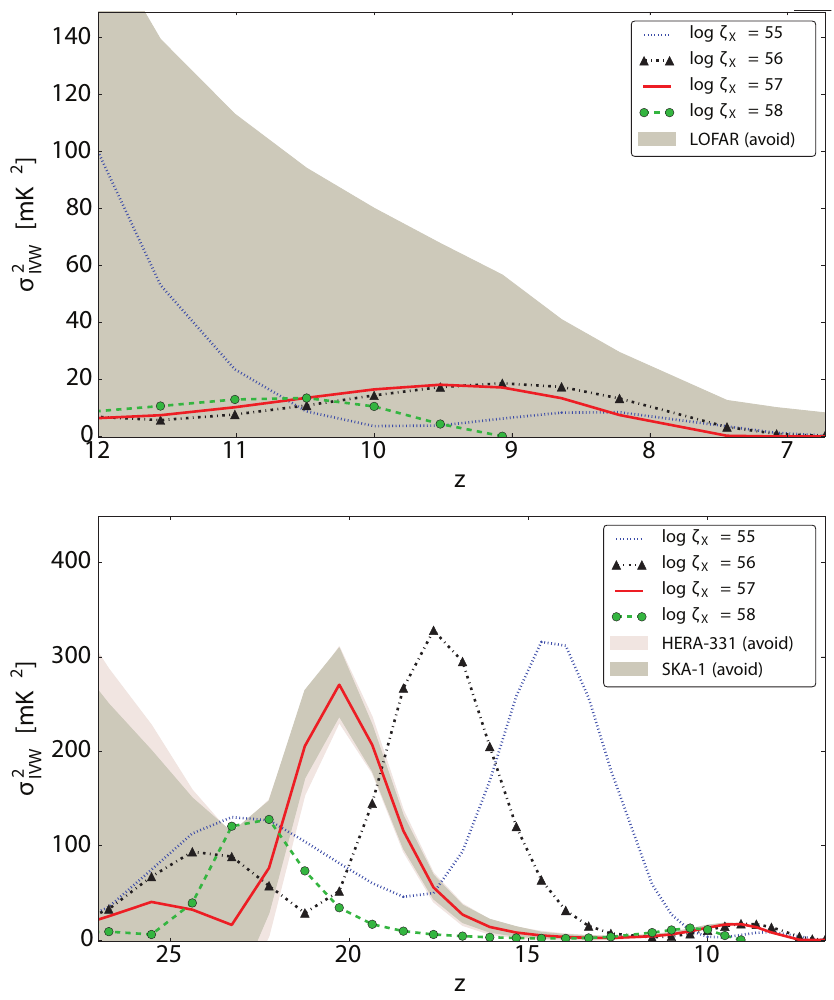}\\
  \caption{Inverse-variance weighted variance as measured from the foreground-free EoR window (i.e. by assigning infinite noise to power from foreground corrupted modes) by LOFAR (top), as well as by HERA and SKA (bottom).
In this worst case scenario, LOFAR will be limited to placing upper bounds on the EoR, while
HERA and SKA could still tightly constrain the EoR. HERA and SKA will still
be able to constrain X-ray heating, but not WF-coupling.}\label{fig:loaf_ivw}
\end{figure}

Although the effectiveness of foreground removal has
yet to be proved (for example, the impact of the 
frequency dependent nature of the instrument 
on the effectiveness of these removal techniques has
yet to be established), we consider optimistically
that it will be possible to remove foregrounds
(described by `remove' 
in the plots of IVW-brightness-temperature variance 
as a function of redshift
in Fig. \ref{fig:simCardVariance}),
and so reduce the wedge's extent to the edges of the instrument's
field of view.
In considering foreground avoidance
(described as `avoid' in the plots of Fig. \ref{fig:loaf_ivw}),
we assume that the spectral structure of the foregrounds only 
extend by $\Delta k_\parallel = 0.1h\,$Mpc$^{-1}$
beyond the wedge described by Equation \ref{eqn:hor_limit}
(in line with the predictions of \citealt{Parsons2012}).
For both foreground
models we assume that baselines sampling
the same $k_\perp$ can be combined coherently.

\begin{table}
    \caption{Summary of the instrumental properties we assume in 
calculating our errors. We assume a bandwidth of $8\,$MHz
and an integration time of $1000\,$hours
for all instruments.
We use exact station locations for MWA and LOFAR
from \citet{Beardsley2012a} and
\citet{VanHaarlem2013} respectively
(we also assume that each LOFAR HBA substation can 
be correlated independently to maximize redundant baselines).
For HERA we assume 331 antenna in a closely packed hexagon
as per \citet{Pober2013a}.
Following \citet{Pober2013a} and \citet{Greig2015a}
we model SKA as a closely packed hexagon (to maximize
redundancy) of 218 antenna
out to $\sim 300$\,m with a further 215 stations randomly 
distributed in an annulus from to $300$-$600$\,m,
217 randomly placed stations in another annulus from 600-1000\,m, 
and 216 randomly placed in an annulus
covering 1000-2000\,m from the centre of the array.}
    \begin{center}
      \begin{tabular}{lcccc}
	\hline 
        \parbox{2.5cm}{Parameter} & LOFAR & MWA & HERA & SKA-1\footnote{We choose to account for the (recently announced) halving of SKA phase 1 collecting area by
reducing the element size rather than the number of 
stations.}\\ 
        \hline      
        \parbox{2.5cm}{Number of stations} & 48 & 128 & 331 & 866\\
        \parbox{2.5cm}{Element size [m]} & 30.8 & 6 & 14 & 35/$\sqrt{2}$\\
        \parbox{2.5cm}{Collecting area [m$^2$]} & 35,762 & 896 & 50,953 & 416,595\\
        \parbox{2.5cm}{Receiver T [K]} & 140 & 440 & 100 & 40\\
        \hline
      \end{tabular}
    \end{center}
    \label{tbl:inst}
\end{table}
We perform an IVW summation over the power spectra
measured by an instrument (using
errors from 21\textsc{cmsense} and the instrumental
properties described in table \ref{tbl:inst})
to get constraints on the brightness-temperature
variance.\footnote{In performing an inverse-variance-weighted
summation over the power spectrum to 
calculate the IVW brightness-temperature variance
we do not worry about normalisation factors as the 
power spectrum as a function of $k$ is not bounded. 
As such, care must be taken
in comparing the amplitude of such an observation with simulations, 
i.e. the IVW variance must be simulated with equivalent
modes to those probed by observations.}
The 1-$\sigma$ error on the IVW variance 
is estimated by
$[\sum_i (1/\delta \Delta^2_i)^2]^{-\frac{1}{2}}$.
An inverse-variance-weighted
sum over the dimensionless power spectrum ($\Delta^2_i$) at different $k_i$
is only an unbiased estimator if the 
noise is Gaussian distributed with zero mean
and the power is approximately
flat (i.e. $\Delta^2_i = \Delta^2$ for all $i$)
so that
$\langle \sum_i (\Delta^2_i+n_i)(1/\delta \Delta^2_i)^2/\sum_i (1/\delta \Delta^2_i)^2 \rangle = \Delta^2$.
Of course this is not strictly true as there
is important evolution in the shape of
the power spectrum with $k$. 
As such calculating 
$\sigma^2_{\mathrm{ivw}} = \sum_i \Delta^2_i(1/\delta \Delta^2_i)^2/\sum_i (1/\delta \Delta^2_i)^2$
means that $\sigma^2_{\mathrm{ivw}}$ is sensitive to the
details of the noise, which is in
turn sensitive to
the details of the instrument.\footnote{Under the assumption
that  $\Delta^2_i = \Delta^2$ for all $i$, then $\sum_i \Delta^2_i(1/\delta \Delta^2_i)^2/\sum_i (1/\delta \Delta^2_i)^2  = \Delta^2$.} 
The limited
resolution of the instruments means that
the power contribution from large $k$ is
totally suppressed; it is this that recovers
the characteristics of the variance seen in
our $\sigma^2_{\mathrm{ivw}}$ statistic. 
Similarly, the IVW-variance is sensitive to
the foreground model we assume; the presence
of a foreground-corrupted wedge means that power from 
the associated $k$ modes will be suppressed in calculating
$\sigma^2_{\mathrm{ivw}}$.
As is clear from the differing amplitudes
between the 
plots of Fig. \ref{fig:simCardVariance} and those
of Fig. \ref{fig:loaf_ivw},
the level of foreground corruption 
can
seriously impact the amplitude of
the variance. We must therefore be very
careful when interpreting this statistic quantitatively
from observations. The power spectra measured
from $k$-modes inside the EoR window should
not suffer from this issue. 
The qualitative nature of the
variance's evolution is insensitive to 
the foreground corruption we consider here
and could therefore be useful
for constraining coupling, X-ray
heating and reionization.

We find that the first-generation instruments
will only be able to constrain our models using the variance
if foreground removal is possible.
If so, then as is clear from the
top row of plots in Fig. \ref{fig:simCardVariance},
both LOFAR and MWA will be able to
constrain reionization and would also be sensitive
to late X-ray heating. 
However, using foreground avoidance
LOFAR could be sensitive
to models in which reionization ends later
than our models assume, but will more
likely be limited to setting upper limits
(see the top plot of Fig. \ref{fig:loaf_ivw}).
Note that because of the maximal redundancy of its
baselines, the next phase of PAPER (consisting of 128-elements,
see \citealt{Ali2015})
is only marginally less sensitive than MWA (see the 
appendix of \citealt{Pober2013}) despite having
less than half the collecting area.

We again emphasise that our fiducial
reionization model is optimistic, and
so first-generation instruments may struggle
more than is suggested by our analysis.
Furthermore, due to the presence of sinks in the IGM 
the variance may be up to a factor
of two smaller than the models of this paper predict.
If extreme levels of remnant \hi in galaxies are present then 
the variance will be reduced even further \citep{Watkinson2015}.
Such reduction of the variance could make it
difficult for first-generation instruments
to do more than place upper limits,
even if foregrounds can be removed.
However, the IVW-variance will inevitably have smaller
errors than the power spectrum at a given 
$k$ and therefore, first-generation instruments
would do well to exploit it in their
quest to make a first detection.

Under the same assumptions, next-generation instruments such as SKA and HERA 
will be able to tightly constrain the variance for the coupling, heating
and reionization epochs (as seen
in the bottom plots of Fig. \ref{fig:simCardVariance}). 
Note that the IVW variance exhibits three distinct
peaks corresponding to WF coupling,
X-ray heating, and reionization;
it does not exhibit the plateau between WF coupling
and X-ray heating as seen in the standard variance 
(i.e. that measured
from a clean simulated box).
Even if foreground removal proves intractable
then the foreground-avoidance technique will 
return strong constraints
on the heating and reionization epochs
(see the bottom plot of Fig. \ref{fig:loaf_ivw}).
Sinks in the IGM will not stop these next-generation
instruments from returning strong constraints
on the EoR using the moments.
However, extreme levels of residual \hi
can qualitatively alter the evolution of the moments from that
described in this paper \citep{Watkinson2015}.

\begin{figure}
  \centering
  \includegraphics{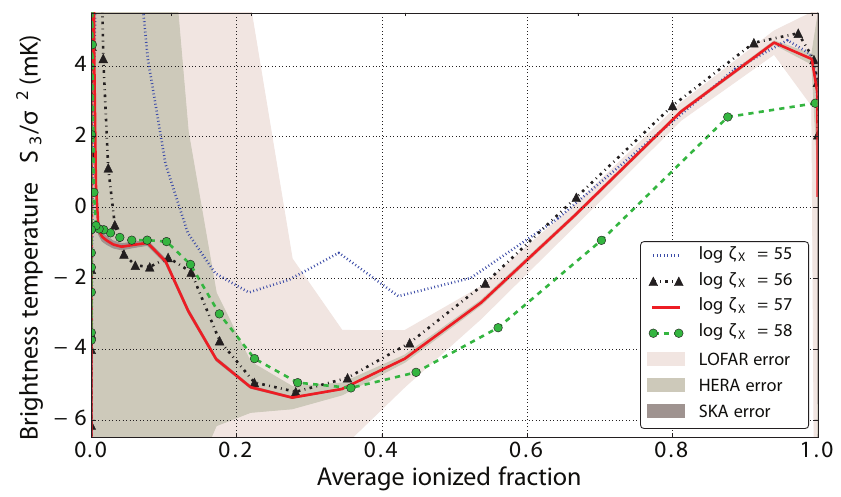}\\
  \includegraphics[trim=1.05cm 0.0cm 0cm 2cm, clip=true, scale=0.25]{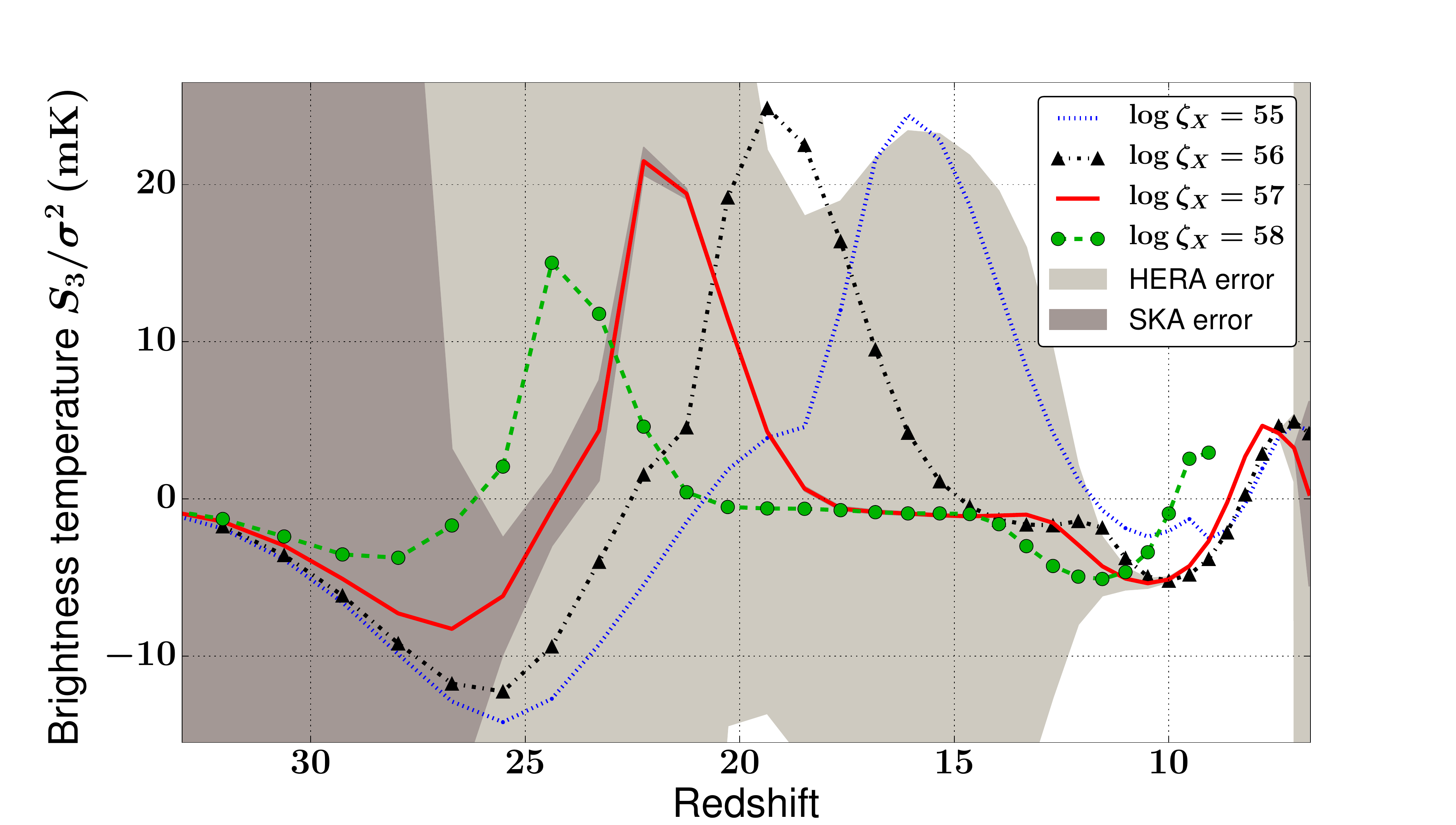}\\
  \caption{Dimensional skewness as a function of ionized fraction (top) and redshift (bottom) from brightness-temperature maps smoothed to a radius of 10\,Mpc (top). Errors (shaded from dark to light for SKA, HERA and LOFAR respectively) correspond to approximate instrumental errors only and are therefore optimistic estimates. Whilst the first-generation instruments will struggle to extract useful constraints from this statistic, the skewness measurements from HERA and SKA will return vital constraints on the EoR. When we examine this statistic as a function of redshift (bottom), we find that SKA will also be able to tightly constrain X-ray heating with the skewness, but that HERA will not be sensitive to the skewness beyond the EoR.}\label{fig:skew_err}
\end{figure}

We use the approach detailed in \citet{Watkinson2014}
to approximate instrumental errors on the skewness,
this approach assumes that foregrounds
can be perfectly removed and approximates
instrumentals by smoothing and re-sampling pixels 
to match the resolution of the telescope.
We plot the dimensional skewness
as a function of both ionized fraction
and redshift in Fig. \ref{fig:skew_err}.\footnote{Prior
to measuring the skewness
for this figure, we smoothed brightness-temperature
boxes to a radius of 10\,Mpc to suppress noise corruption.}
These errors should be viewed as optimistic estimates
and will likely be quite a bit larger.
As an illustration, if we compare the errors on
the variance as calculated by \citet{Watkinson2014}
with those predicted for the IVW variance,
we find its S/N is a factor of order 3
worse if foregrounds can be removed;
if foreground avoidance is necessary
then S/N can be 20 - 50 times worse.

We see that it will be possible to use the 
skewness to constrain models of late X-ray heating,
possibly with LOFAR but certainly with the
next-generation instruments. 
Therefore this presents an excellent
opportunity for these telescopes to constrain
a fundamental property of the Universe's evolution, 
namely the relative timing of WF coupling, 
X-ray heating, and reionization. 

\section{Discussion} \label{sec:Disc}

There are several approximations
made in \cmfast
that may have important repercussions;
in particular the code assumes 
average properties of the IGM in calculating the
X-ray mean free path. This is
most important during the later
stages of reionization when large ionized
regions will result in fluctuations 
of the X-ray mean free path between sight-lines
in the box. 
The code also assumes either population II or
III stellar spectra in its calculations
of WF coupling, and does not account
for the possibility of mixed populations,
feedback effects or shot noise. 
The nature of these first stars 
(and of the remnants they leave when 
they die) is very uncertain. 
Recent simulations indicate that these first 
stars will be $10-100\,M_{\odot}$ and are expected to form 
in small clusters (e.g. \citealt{Hirano2014, Greif2011}).
Formation of such population III 
stars rely primarily on cooling via molecular hydrogen,
however they produce large amounts
of Lyman-Werner radiation (which disassociate molecular hydrogen)
and so are likely to stunt further formation of population III stars
(e.g. \citealt{Wise2007, OShea2008}).
Such large stars are also short lived, 
so it is not unreasonable 
(as is done in this work) to 
assume that population II stars
will be the dominant
driver of the processes discussed here. 
However, it is possible that these results
are inaccurate during the very early
phases of coupling when the very first
stars form.

Whilst simulations such as \cmfast have been
tested against numerical simulations
during the EoR assuming that
the spin temperature is saturated 
(see for example, \citealt{Zahn2011} and \citealt{Majumdar2014}),
there has not been equivalent tests
of these when spin temperature
fluctuations are included.
This is mainly because numerical 
simulations with the necessary scale
and resolution do not yet exist.
The only numerical simulations that perform radiative-transfer
in all of the relevant frequency bands are those of
\citet{Baek2010}. 
These simulation do not resolve haloes
with mass below 10$^{10}M_{\odot}$,
therefore they do not resolve atomically cooling haloes.
As such, all astrophysical processes are driven
by more massive, and therefore more rare and biased,
haloes than is to be expected in reality.
It is therefore not possible to draw direct comparison 
between \citet{Baek2010} and \textsc{\small 21CMFAST }.
However, \cite{Mesinger2013a} note that
the qualitative evolution of the power
spectrum at $k\sim0.1$ of \cmfast (when including 
spin temperature fluctuations) is in agreement
with the numerical simulations of \citet{Baek2010}.
We also find that the skewness of our late
X-ray heating model ($\zeta_{\textsc{x}}=10^{55}$)
qualitatively agrees with their S6 model,
which is encouraging.

There are other processes that must be considered in 
parallel to spin temperature fluctuations. 
For example, and as already discussed,
the presence of sinks could drastically reduce the variance. 
This reduction is due to residual signal in 
ionized regions and sub-pixel ionized regions. 
X-ray ionizations will occur in a more homogeneous
fashion than UV ionizations and so will be
responsible for partially ionizing regions
outside of UV carved ionized regions. 
It therefore seems likely that the reduction of
variance caused by X-ray ionizations will
be in addition to that caused by sinks, i.e.
they will further reduce the contrast between over 
and under-dense regions. 
The simulation of \cite{Sobacchi2014a} also incorporate 
UVB feedback which suppresses star formation, 
such feedback will clearly impact on both
the Lyman-$\alpha$ and X-ray production.
However, given the large amplitudes
seen in the one-point statistics during the heating epoch,
and that we have studied four orders of magnitude in the X-ray
efficiency, it is unlikely that UVB feedback will
have a dramatic effect beyond that seen here.

These examples (and the lack of numerical
simulations with which to test \textsc{\small 21CMFAST })
serve to illustrate the
challenge we face in
simulating the epochs of the first
dawn and reionization.
The results of this work
should therefore not be considered conclusive
and it is essential that we do more to understand
how the statistics of the 21-cm moments are
impacted by different physical processes (and their interplay).

\section{Conclusion} \label{sec:Conc}
In this paper, we have considered the sensitivity
of one-point statistics of the 21-cm brightness
temperature to fluctuations in WF coupling
and X-ray heating,
concentrating on the skewness and
the variance. 
We use semi-numerical simulations to
vary the efficiency at which X-rays are produced
(to cover four orders of magnitude) and the spectral index
of the X-ray SED (to encompass the range of observational
constraints we have at low redshifts).
From this study we establish that:
\begin{packed_enum}
\item the location and amplitude of the global maxima
in the redshift evolution of both the skewness
and variance are
sensitive to the X-ray production efficiency.
The amplitude of this maximum in
the variance is also sensitive to the 
hardness of the X-ray SED.
This degeneracy may be broken,
as the skewness is only sensitive to the
X-ray production efficiency;
\item late X-ray heating causes the
drop from the X-ray heating peak to occur at
an ionized fraction of about a quarter
rather than in the very early stages of reionization.
In such a model, the turnover in the variance,
usually associated with the mid-point of
reionization, is shifted to higher
ionized fractions.
The evolution of the skewness is qualitatively
different if X-ray heating occurs late,
this provides a clean way to constrain such a model.
The amplitude of the
variance is greatly reduced in these models,
which would make it more challenging
for the first-generation instruments 
(such as LOFAR, MWA and PAPER) to make a detection
of reionization using the variance;
\item the high-redshift heating peak
must be allowed for in models used for parameter
estimation from one-point statistics.
If not our inferences may be very misleading.
This is equally true for performing parameter estimation
from the power spectrum;
\item X-ray ionizations reduce the amplitude
of the variance.
In most models we consider
they reduce the variance
by $\sim 10\%$ during the mid-phase of reionization;
in the most X-ray efficient model, we find this
reduction to be $\sim 25\%$.
\end{packed_enum} 

We consider (for the first time to the authors' knowledge)
the variance as measured using foreground avoidance techniques.
From this we find that the next-generation instruments such
as HERA and SKA will return strong constraints on both
reionization and X-ray heating, even if we are unable 
to remove foregrounds.

The findings of this paper will help us to 
correctly interpret future observations of the 21-cm
brightness temperature;
in particular they have important consequences
for improving parameter estimation during reionization.

\section*{Acknowledgements}
We thank Andrei Mesinger for making the {\small 21CMFAST} code 
used in this paper publicly available as well as for useful comments. 
CW is supported by an 
STFC studentship. JRP acknowledges support under FP7-PEOPLE-2012-CIG 
grant \#321933-21ALPHA and STFC consolidated grant ST/K001051/1.

\bibliographystyle{mn2e}


\appendix
\section{Analytical expression for the skew of $\Delta\mu$}\label{sec:appen}
\begin{equation}
\begin{split}
S_{3,\Delta\mu} &= 3 \left\langle \delta \left(\Trat\right)^2 \right\rangle 
-3\left\langle \delta^2\Trat \right\rangle
+6\left\langle\delta\left(\Trat\right)^2 \right\rangle \\
&-\left\langle \left(\Trat\right)^3\right\rangle
-3\left\langle \delta\left(\Trat\right)^3\right\rangle
-3\left\langle \delta^2 \left(\Trat\right)^3\right\rangle \\
&+\left\langle \delta^3 \right\rangle
-3\left\langle \delta^3\Trat\right\rangle
+3\left\langle \delta^3 \left(\Trat\right)^2\right\rangle \\
&-\left\langle \left(\delta\Trat\right)^3 \right\rangle
-6\left\langle \Trat\right\rangle\left\langle \delta\Trat\right\rangle \\
&+3\left\langle \Trat\right\rangle \left\langle \left(\Trat\right)^2 \right\rangle + 6\left\langle \Trat\right\rangle\left\langle\delta\left(\Trat\right)^2 \right\rangle \\
&+3\left\langle \Trat\right\rangle \left\langle \delta^2\right\rangle
-6\left\langle \Trat \right\rangle \left\langle \delta^2\Trat\right\rangle \\
&+3\left\langle \Trat\right\rangle\left\langle \left(\delta\Trat\right)^2\right\rangle
-6\left\langle \delta\Trat\right\rangle^2\\
&+3\left\langle \left(\Trat\right)^2\right\rangle\left\langle \delta\Trat\right\rangle
+6\left\langle \delta\Trat\right\rangle\left\langle\delta\left(\Trat\right)^2 \right\rangle \\
&+3\left\langle \delta\Trat\right\rangle\left\langle \delta^2\right\rangle
-6\left\langle \delta\Trat\right\rangle\left\langle \delta^2\Trat \right\rangle\\
&+3\left\langle\delta\Trat \right\rangle\left\langle\left(\delta\Trat\right)^2 \right\rangle
-2\left\langle \Trat\right\rangle^3\\
&-6\left\langle \Trat\right\rangle^2\left\langle \delta\Trat\right\rangle
-6\left\langle \Trat\right\rangle \left\langle \delta\Trat\right\rangle^2\\
&-2\left\langle \delta \Trat\right\rangle^3
\,,\\
\end{split}\label{eqn:anal_vari}
\end{equation}

\bsp
\end{document}